\documentclass[preprint, prd, 12pt]{revtex4-2}
\usepackage[normalem]{ulem}
\usepackage[brazil, english]{babel}
\usepackage[utf8]{inputenc}
\usepackage{graphicx}
\usepackage{epsfig}
\usepackage{placeins}
\usepackage{amssymb}
\usepackage{amsmath} 
\usepackage[unicode=true,bookmarks=false,breaklinks=false,pdfborder={0 0 1},colorlinks=true]
 {hyperref}
\hypersetup{citecolor=blue,linkcolor=magenta,urlcolor=red}

\newcommand{\bee}{\begin{equation}}
\newcommand{\eee}{\end{equation}}
\newcommand{\eaa}{\end{eqnarray}}
\newcommand{\baa}{\begin{eqnarray}}

\def\ni{\noindent}

\def\p{\partial}
\def\r+{r_{\!\!{+}}}

\begin{document}

\title{Joule-Thomson expansion for quantum corrected AdS-Reissner-Nordström black holes in Kiselev spacetime with Barrow fractal entropy}

\author{Everton M. C. Abreu$^{1,2}$}
\email[Eletronic address: ]{evertonabreu@ufrrj.br}
\author{Henrique Boschi-Filho$^{3}$}
\email[Eletronic address: ]{boschi@if.ufrj.br}  
\author{Rafael A. Costa-Silva$^{3}$}
\email[Eletronic address: ]{rafaelcosta@pos.if.ufrj.br}
\affiliation{
$^{1}$Physics Department, Universidade Federal Rural do Rio de Janeiro,  RJ, Brazil.\\
$^{2}$Applied Physics Graduate Program, Physics Institute, Universidade Federal do Rio de Janeiro, RJ, Brazil. \\
$^3$ Instituto de F\'isica, Universidade Federal do Rio de Janeiro, RJ, Brazil. 
 }

\date{\today}

\begin{abstract}
 Recently, Barrow proposed an extension of the Bekenstein-Hawking black hole entropy to include the effects of fractal geometry through a parameter $\Delta$. 
 Since then, several interesting issues related to this extended entropy have been explored in the literature. In this work, we investigate the effects of the fractal parameter $\Delta$ on the inversion temperature connected to the Joule-Thomson expansion that can be obtained from the thermodynamics of AdS-Reissner-Nordström black holes in Kiselev spacetime. The description of such physical systems involves numerical solutions concerning the fractal parameter $\Delta$. The results are shown by temperature-pressure curves for multiple values of the set of parameters that define the black hole thermodynamics. In conclusion of our analysis, we also show isenthalpic curves corresponding to fixed-mass black hole processes.
\end{abstract}

\maketitle
\newpage


\section{Introduction}

Since black holes (BHs) are considered thermodynamic systems through the works of Bekenstein and Hawking (BeH) \cite{Bekenstein:1973ur,Hawking:1974sw}, they are the targets of a very intense study
(see, {\sl e.g.,} 
\cite{Mann:2025xrb, Wald:2025nbz, Elizalde:2025iku, Witten:2024upt, Ruppeiner:2023wkq, Ong:2022frf, Wei:2022dzw, Auffinger:2022khh, Bravo-Gaete:2022lno, Eisert:2008ur, Berti:2009kk}). Besides their thermodynamic properties, BHs are the most intrigate objects in the cosmic structure because they have strong gravitational forces.   Hence, we can recognize the associations between the laws of general relativity, thermodynamics, and quantum mechanics. The direct consequence is that BHs can be used to fathom quantum gravity. BHs have a so-called event horizon where, due to strong gravitational attraction, nothing passes through it to escape from the BH accretion disk which is a framework formed by materials such as dust and gas. However, S. Hawking proposed that through a quantum mechanical process BHs produce radiation, the so-called Hawking radiation. 

In 2020, based on the shape of the COVID-19 virus, J. Barrow \cite{barrow} suggested a modification of the BeH entropy 
\begin{equation}
    S= \Big( \frac{A}{4} \Big)^{1+\Delta/2}\,, 
    \label{S_Barrow}
\end{equation}
where $A$ is the area of the BH horizon and $\Delta$ is a parameter related to its fractality. 
Specifically, he was motivated by the kind of Koch snowflake shape.   Hence, he introduced an exponent, in BeH expression, formed by the $\Delta$-parameter. A parameter that varies from zero to one $(0 \leq \Delta \leq 1)$ where $\Delta=0$ means the original expression of BH entropy and $\Delta=1$ means that the event horizon has the most intricate geometry, the fractal one.   Since then, the Barrow entropy formulation has been used to investigate several interesting and relevant issues in theoretical physics (see {\sl e.g.} \cite{Abdalla:2022yfr, Saridakis:2020zol, Anagnostopoulos:2020ctz, Nojiri:2022aof, Barrow:2020kug, Dimakis:2021gby, Moradpour:2020dfm, Sheykhi:2021fwh, Nojiri:2021czz, Abreu:2020wbz}). 
In addition, the $\Delta$-parameter was also considered not a constant but a variable parameter \cite{Abreu:2024uvp, Abreu:2025etn, Basilakos:2023seo}.

On the other hand, the conjecture of a variable cosmological constant has consequences such as heat cycles, phase transitions, and compressibility of BHs \cite{dolan, Johnson:2014yja, Caceres:2015vsa, Johnson:2015ekr, Johnson:2015fva, Hendi:2017bys, Johnson:2016pfa, Chandrasekhar:2016lbd, Wei:2017vqs}. Taking into account anti-de Sitter (AdS) BHs 
\cite{Maldacena:1997re, Witten:1998zw, Chamblin:1999tk, Chamblin:1999hg, Kastor:2009wy, Kastor:2009wy, Kubiznak:2012wp, Lobo:2022eyr}, these thermodynamic effects motivated several authors \cite{Okcu:2016tgt, Okcu:2017qgo, Mo:2018rgq, Lan:2018nnp, Mo:2018qkt, Cisterna:2018jqg, Rizwan:2018mpy, Chabab:2018zix, Yekta:2019wmt, Li:2019jcd, Nam:2019zyk, K.:2020rzl, Bi:2020vcg, boschiJT1, boschiJT2, Silva:2021qtw, Yin:2021akt, Biswas:2021uop, Xing:2021gpn, Cao:2022hmd, Mondal:2022ozv, Bai:2022yno, Masmar:2023qol, Li:2023mql, Alipour:2024xxs, Yasir:2024gza, Javed:2024ohv,  Dai:2025fxl} to consider the Joule-Thomson expansion (JTE) in various gravity contexts. In other words, in JTE, a gas at high pressure passes through a porous wall to another part of the recipient at low pressure (see {\sl e.g.}  \cite{Reif}). Because the experiment is carried out under adiabatic conditions, the enthalpy is preserved, i.e., it is constant. To discuss the JTE means that we can consider the heating-cooling effect.

Recently, some authors have investigated the JTE considering Barrow's modified BH entropy  in different gravity scenarios:  with nonlinear electrodynamics \cite{Javed:2024nnt};  F(R) gravity \cite{Yasir:2024oah}; 
 Rastall gravity \cite{Javed:2025dit};   spacetime foam \cite{Ladghami:2024yjn}; and Einstein-Maxwell-Scalar gravity \cite{Biswas:2025jxb}. 

The main objective of this paper is to investigate the JT effect modified by the Barrow entropy on quantum corrections and cosmological fluids described by Kiselev spacetime, such as phantom dark energy and quintessence, in an  AdS-Reissner-Nordström (AdS-RN) BH. An investigation of JTE in a quantum corrected AdS-Reissner-Nordström black hole in Kiselev spacetime with standard BeH entropy was presented in \cite{boschiJT1}, so the present work can be considered as an extension to study the effects of the fractal Barrow entropy on JTE isenthalpics and inversion temperature in these BHs. 

Quantum corrections are interesting relative to the phenomenology of microscopic BHs. Consequently, the effects in a static BH have been analyzed by Kazakov and Solodukhin \cite{Kazakov:1993ha}, where the authors considered small deformations in the Schwarzschild metric due to quantum fluctuations in the gravitational and matter fields. The effect of quantum corrections on phase transitions was described in Ref. \cite{Lobo:2019put}.

Furthermore, the consideration of a cosmological fluid is important because it is present both in the current and in the early universe.  To incorporate cosmological fluids, we will use here the Kiselev metric \cite{Kiselev:2002dx}, where such matter can be depicted through an EoS like $P = \, \omega \, \rho$. For example, when $\omega = -2/3 $, it can be affirmed that the environment around the BH is a kind of quintessence. Besides, when $\omega < - 1$ the environment is a kind of phantom dark energy \cite{Caldwell:1999ew} (see also \cite{Visser:2019brz, Shahjalal:2019pqb, Lobo:2019put}).

This work is organized as follows: in Section \ref{Sec:JT}, we give a brief review of the JTE. In Section \ref{Sec:Barrow} we computed analytically all the relevant objects concerning the JTE within the Kiselev spacetime with the Barrow entropy. In Section \ref{Sec:Num} we show the numerical results for the JTE and isenthalpyc curves. Finally, in section \ref{Sec:Conc} we present our conclusions and our final observations.


\section{Joule-Thomson expansion: a brief review}\label{Sec:JT}

In JTE, as we mentioned before, a high-pressure gas propagates through a porous wall or a valve to another part of the recipient with low pressure in an adiabatic tube, and the enthalpy continues to be constant throughout the expansion process (for a review of JTE in van der Waals gasses and an application to AdS-RN BH, see \cite{Okcu:2016tgt})

We can consider the variation of the temperature $T$ relative to the pressure $P$, at constant enthalpy $H$, and this change is given by \cite{Reif}
\begin{equation}
    \label{muJT}
    \mu_{\rm JT}\,=\,\Big(\frac{\partial T}{\partial P}\Big)_H\,,
\end{equation}

\noindent which is called the JT coefficient. The sign of $\mu_{\rm JT}$ rules the cooling or heating. As pressure decreases, the variation of pressure is negative, but the variation of temperature may be negative or positive.   If the temperature variation is negative (positive), then the JT coefficient is positive (negative).   Hence, the gas cools (warms).

From the first law of thermodynamics,
\begin{equation}
    \label{2}
    dU\,=\, TdS - PdV \,\,,
\end{equation}

\noindent the enthalpy is well known as
\begin{equation}
    \label{3}
    H\,=\,U + PV
\end{equation}

\noindent and its variation is given by
\begin{equation}
    \label{4}
    dH\,=\,TdS + VdP
\end{equation}

But, as we said above, $dH=0$, hence from Eq. \eqref{4}
\begin{equation}
    \label{5}
    T\Big(\frac{\partial S}{\partial P}\Big)_H\,+\,V\,=\,0\,\,.
\end{equation}

The entropy is a state function, consequently,
\begin{eqnarray}
    \label{6}
&& dS\,=\,\Big(\frac{\partial S}{\partial P}\Big)_T\,dP\,+\,\Big(\frac{\partial S}{\partial T}\Big)_P\,dT \nonumber \\
&& \Longrightarrow \Big(\frac{\partial S}{\partial P}\Big)_H\,=\, \Big(\frac{\partial S}{\partial P}\Big)_T \,+\, \Big(\frac{\partial S}{\partial T}\Big)_P\,\Big(\frac{\partial T}{\partial P}\Big)_H
\end{eqnarray}

Substituting Eq. \eqref{6} into Eq. \eqref{5}, we have the following.
\begin{equation}
    \label{7}
    T\,\Bigg[\Big(\frac{\partial S}{\partial P}\Big)_T \,+\, \Big(\frac{\partial S}{\partial T}\Big)_P\,\Big(\frac{\partial T}{\partial P}\Big)_H \Bigg]\,+\, V \,=\,0 \,\,.
\end{equation}

We will use the Maxwell relation,
\begin{equation}
    \label{8}
    \Big(\frac{\partial S}{\partial P}\Big)_T \,=\,-\,\Big(\frac{\partial V}{\partial T}\Big)_P
\end{equation}

\ni and the heat capacity at constant pressure
\begin{equation}
    \label{9}
    C_P\,=\,T\,\Big(\frac{\partial S}{\partial T}\Big)_P \,\,.
\end{equation}

Substituting Eqs. \eqref{8} and \eqref{9} into Eq. \eqref{7}, one obtains 
\begin{equation}
    \label{10}
    -\,\Big(\frac{\partial V}{\partial T}\Big)_P \,+\, C_P\, \Big(\frac{\partial T}{\partial P}\Big)_H \,+\, V \,=\,0\,, 
\end{equation}

\ni and so the JT coefficient is given by
\begin{equation}
    \label{11}
    \mu_{\rm JT}\,=\, \Big(\frac{\partial T}{\partial P}\Big)_H \,=\, \frac{1}{C_P}\,\Bigg[ T\Big(\frac{\p V}{\p T}\Big)_P\,-\,V \Bigg]\,. 
\end{equation}

From Eq. \eqref{muJT} we see that the JT coefficient measures the temperature variation of a gas or liquid when it expands without carrying out work keeping the enthalpy constant.   At the inversion temperature, there is no cooling or warming and so there is no temperature variation, and the JT coefficient is zero.   At $T_{\rm i}^{\rm JT}$ the intermolecular forces are in equilibrium.   We will have cooling only at $T$ lower than $T_{\rm i}^{\rm JT}$.

The inversion temperature in the Joule-Thomson expansion is characterized by a zero $\mu_{\rm JT}$, so
\begin{equation}
    \label{12}
    T_{\rm i}^{\rm JT} \,=\, V\,\Big(\frac{\p T}{\p V}\Big)_P \,\,.
\end{equation}

This result is very useful for computing the heating and cooling parts in the $T-P$ planes.


\section{Barrow black holes thermodynamics and inversion temperature in Kiselev spacetime}\label{Sec:Barrow}

In this section, we will thoroughly analyze the quantum correction of AdS-RN BH that is embedded in Kiselev spacetime.  Our intention is to carry out a profound analysis of its features.  Specifically,
the thermodynamic topology is studied from the perspectives of generalized JTE, which is very 
important here. We will explore the spacetime metric of the quantum corrected charged AdS BH immersed in a Kiselev spacetime \cite{Kiselev:2002dx, Caldwell:1999ew, Visser:2019brz, Shahjalal:2019pqb, Lobo:2019put, boschiJT1}. 


\subsection{Kiselev black hole}

The AdS-RN metric in Kiselev spacetime, also known as the Kiselev BH, is given by
\bee
\label{metric}
ds^2\,=\, f(r) dt^2 \,-\, \frac{dr^2}{f(r)}\,-\,r^2\,d\Omega^2\,, 
\eee

\ni where the metric $d\Omega^2\,=\, d\theta^2\,+\, \sin^2 \theta d\phi^2$ corresponds to a two-sphere and the horizon function $f(r)$ for a BH with mass $M$, electric charge $Q$, quantum correction $a$, and Kiselev parameters $c$ and $\omega$ is given by
\bee
\label{f(r)}
f(r) \,=\, -\, \frac{2M}{r} \,+\, \frac{\sqrt{r^2-a^2}}{r}\,+\, \frac{r^2}{l^2}\,+\, \frac{Q^2}{r^2} \,-\,\frac{c}{r^{3\omega+1}}\,. 
\eee

\ni Note that $r > a$ is required to avoid the generation of imaginary structures. The quantity $M$ denotes the BH mass, while the symbol $l$ measures the length scale relevant to the asymptotically AdS spacetime. The parameter $c$ is connected to the cosmological fluid comprising the BH, and $Q$ indicates the electric charge of the BH.  It is important to note at this point that from Eq. \eqref{f(r)}, the parameter $a$ is a quantum correction included in the metric that alters, deforms, the fabric of the spacetime.  We can clearly see that it shifts the singularity point, the curvature, the horizon, and the causal structure of the system. The parameter $a$ is associated with the modifications in BH mass due to quantum corrections. The fundamental theory relative to this parameter is deeply debated in \cite{Kazakov:1993ha, Shahjalal:2019pqb}. As an independent object, $a$ has the distinctive feature that, when it is zero, the metric reverts to the well known AdS-RN metric, now involved in a cosmic fluid. Supposedly, $a$ can have any value as long as it is smaller than the event horizon’s radius, coherently related to the notion that $a$ constitutes a minor modification to the conventional BH metric.

To begin our analysis, it is very important to understand why we chose this specific metric. M. Visser \cite{Visser:2019brz} has proposed that the Kiselev BH model can be augmented to include a spacetime with $N$ components. The extension is marked by a linear correlation between pressure and energy for each component, as detailed in \cite{Kiselev:2002dx, Visser:2019brz}. In our analysis, we consider various values for $\omega$, such
as $\omega =-1/3,$ $\,$ $\omega=-2/3,$ $\,$ $\omega =-1$ and $\omega =-4/3$. These particular values of $\omega$ have different interpretations, for instance, the phanton dark energy for $\omega<-1$, or the presence of quintessence when $\omega=-2/3$.


\subsection{First law, Barrow entropy, and Inversion Temperature}

The first law of thermodynamics can be written as follows.
\begin{equation}
dH=TdS + VdP + \Phi dQ + \mathcal{C}dc + \mathcal{A}da\,,
\end{equation}
 
\ni where the enthalpy $H$ is identified with the BH mass $M=H$, while $\Phi$, $\cal C$, and $\cal A$, respectively, are the potentials of the electric charge, the Kiselev $c$ and the quantum correction parameter $a$.  For the inversion temperature, $T_{\rm i}^{\rm JT}$, it can be shown, as we mentioned in the last section, that $\mu_{\rm JT}=0$.

To calculate the mass of the BH, we can use the metric function in Eq. \eqref{f(r)}, to obtain  $r_+$, which is the highest root of $f(\r+)=0$, representing the radius of the external horizon. Hence,
\bee
\label{Mass}
M\,=\,\frac 12 \Bigg[ \sqrt{\r+^2-a^2} + \frac{\r+^3}{l^2} + \frac{Q^2}{\r+} - \frac{c}{\r+^{3\omega}}\Bigg]\,.
\eee

Then, the Barrow entropy \cite{barrow} can be written as
\begin{equation}
    S= (\pi \r+^2)^{1+\frac{ \Delta}{2}}\,, 
    \label{S_Delta}
\end{equation}

\ni where $\Delta=0$ recover the Bekenstein-Hawking entropy \cite{Bekenstein:1973ur, Hawking:1974sw} and $\Delta=1$ means that we have the most intricate geometry, the fractal one, i.e., $0 \leq \Delta \leq 1$.  Hence, we have a fractal event horizon embedded in the Kiselev spacetime. Now we have a quantum correction manifested in the entropy expression. In this Barrow model, it also deforms the structure of the event horizon as a function of the $\Delta$-parameter. In addition, it also deforms the way information is stored in the horizon, creating a quantum microstructure or a fractal one.  

Recall that the temperature $T$ can be written as
 \begin{equation}
     T=\bigg(\frac{\partial H}{\partial S}\bigg)_{P,Q}=\bigg(\frac{\partial M}{\partial S}\bigg)_{P,Q}=\bigg(\frac{\partial M}{\partial \r+}\frac{\partial \r+}{\partial S}\bigg)_{P,Q}, 
     \label{T}
 \end{equation}
and that pressure $P$ is related to AdS radius $l$ by \cite{Kubiznak:2012wp}:
\bee\label{Pl}
P =\frac {3}{8\pi l^2}.
\eee

Hence, we can write the BH temperature as
 \begin{equation}\label{temp}
     T=\frac{1}{2(2+\Delta)(\sqrt{\pi} \r+)^{\Delta}}\bigg(8 P\r+-\frac{Q^2}{\pi \r+^3}+\frac{1}{\pi\sqrt{\r+^2-a^2}}+\frac{3\omega c}{\pi \r+^{2+3\omega}}\bigg)\,,  
 \end{equation}
which reduces to the Hawking temperature of the AdS-Reissner-Nordström BH in Kiselev spacetime with quantum corrections in the limit $\Delta\to 0$. 
In addition, the Joule-Thomson inversion temperature is 
 \begin{equation}
     T_{\rm i}^{\rm JT}=V\bigg(\frac{\partial T}{\partial V}\bigg)_{P}\,\,,
 \end{equation}

\ni and the volume of BH 
 \begin{equation}
     V=\frac{4\pi \r+^3}{3},
 \end{equation}

\ni so that we can write
\begin{equation}
    T_{\rm i}^{\rm JT}=\frac{4\pi \r+^3}{3}\bigg(\frac{\partial T}{\partial \r+}\frac{\partial \r+}{\partial V}\bigg)_{P}.
\end{equation}

After some algebraic work, we have
\begin{equation}\label{inversiontemp}
    T_{\rm i}^{\rm JT}= \frac{1}{6(2+\Delta)(\sqrt{\pi} \r+)^{\Delta}}\bigg[8P_{i}^{\rm JT}(1-\Delta)\r++\frac{Q^2(3+\Delta)}{\pi \r+^3}-\frac{3\omega c(2+\Delta+3\omega)}{\pi \r+^{2+3\omega}}+\frac{\Delta a^2 -(1+\Delta)\r+^2}{\pi(\r+^2-a^2)^{3/2}}\bigg]\,.
\end{equation}

Using $T=T_{\rm i}^{\rm JT}$, $P=P_i^{\rm JT}$ in Eq. (\ref{temp}),  subtracting this equation from Eq.  (\ref{inversiontemp}), we obtain a relation between $\r+$ and $P_{\rm i}^{\rm JT}$, which is given by
\begin{equation}
  \frac{\r+^3 + \r+ (3+\Delta)(\r+^2-a^2 )}{( \r+^2-a^2 )^{3/2}} 
+ 8 \, P_{\rm i}^{\rm JT} \, \pi \, \r+^2 (2+\Delta) 
- \frac{Q^2 (6+\Delta)}{\r+^2} 
+ 3 \, c \, \r+^{-1-3\omega} \, \omega \, (5  + 3\omega+\Delta) = 0\,. 
\end{equation}

\ni We will solve this equation numerically to find $\r+(P)$. Then, we can substitute this result into Eq. (\ref{inversiontemp}) to construct a $T\times P$ plot where we can see the inversion temperature curve. This will be analyzed in detail in the next section.


\section{Numerical results}\label{Sec:Num}

\begin{figure}[ht!]
\vskip0.5cm 
\centering
\includegraphics[width=7.8cm]{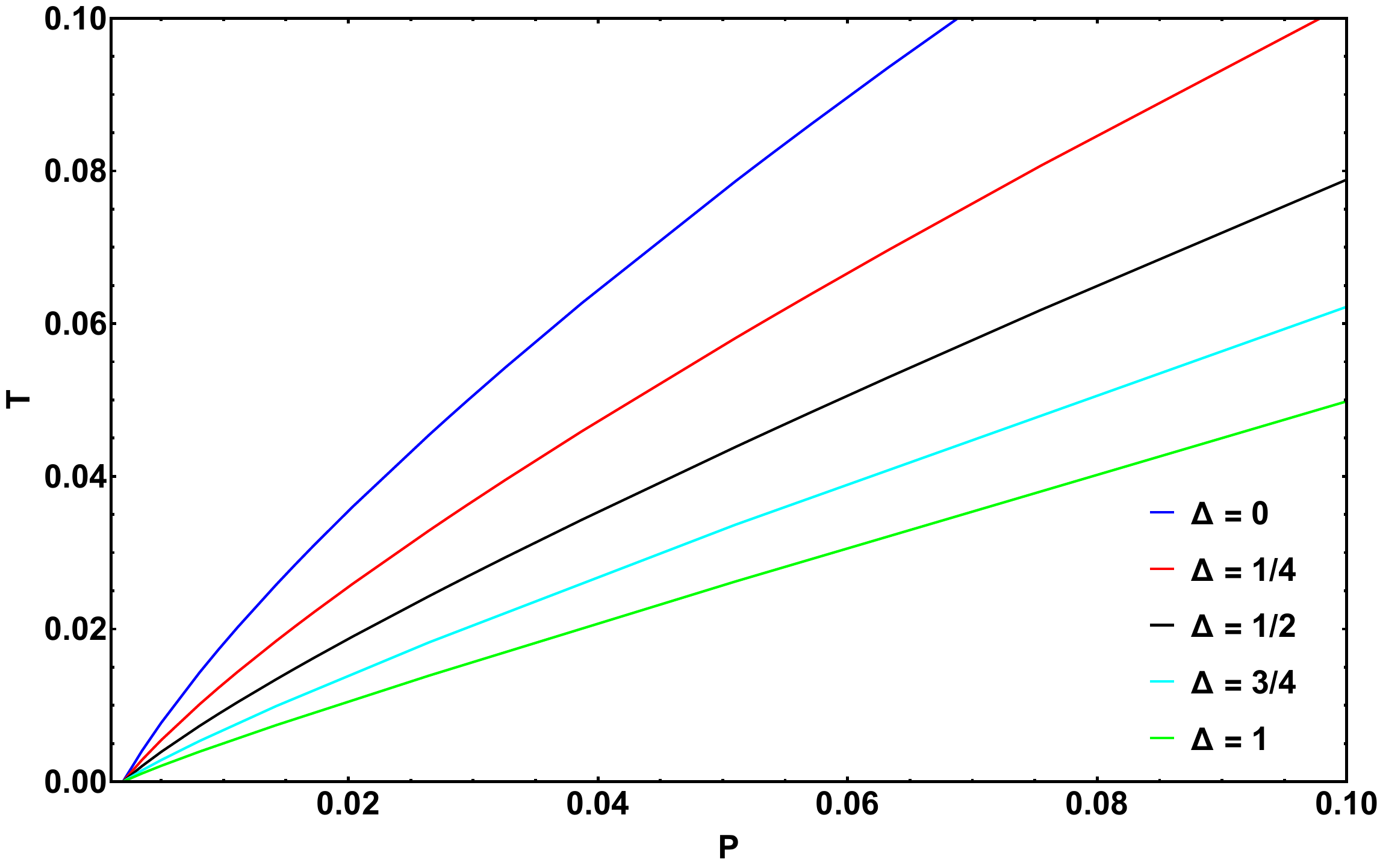}\qquad \includegraphics[width=7.6cm]{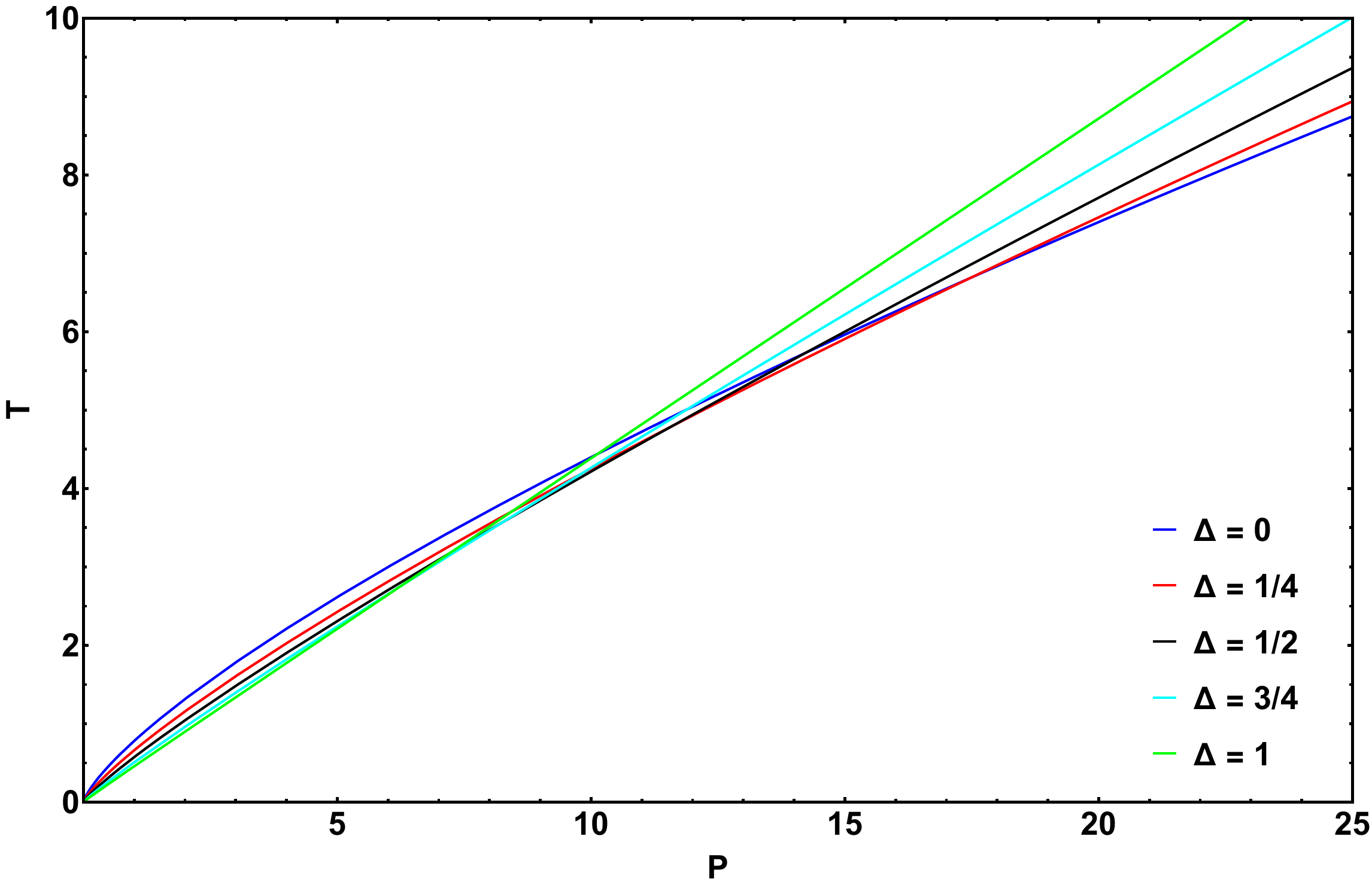}\\
\includegraphics[width=7.8cm]{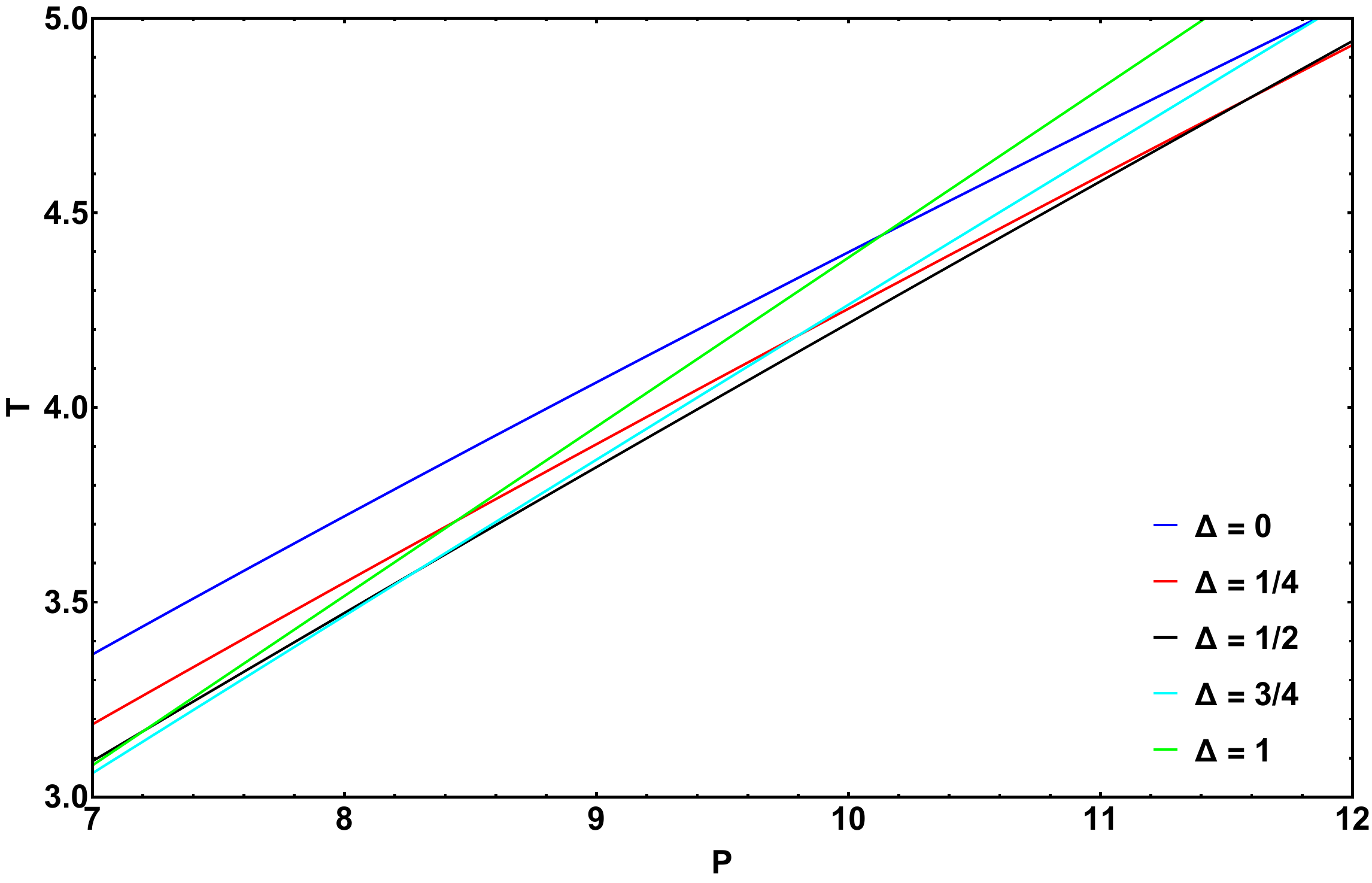}\qquad\includegraphics[width=7.8cm]{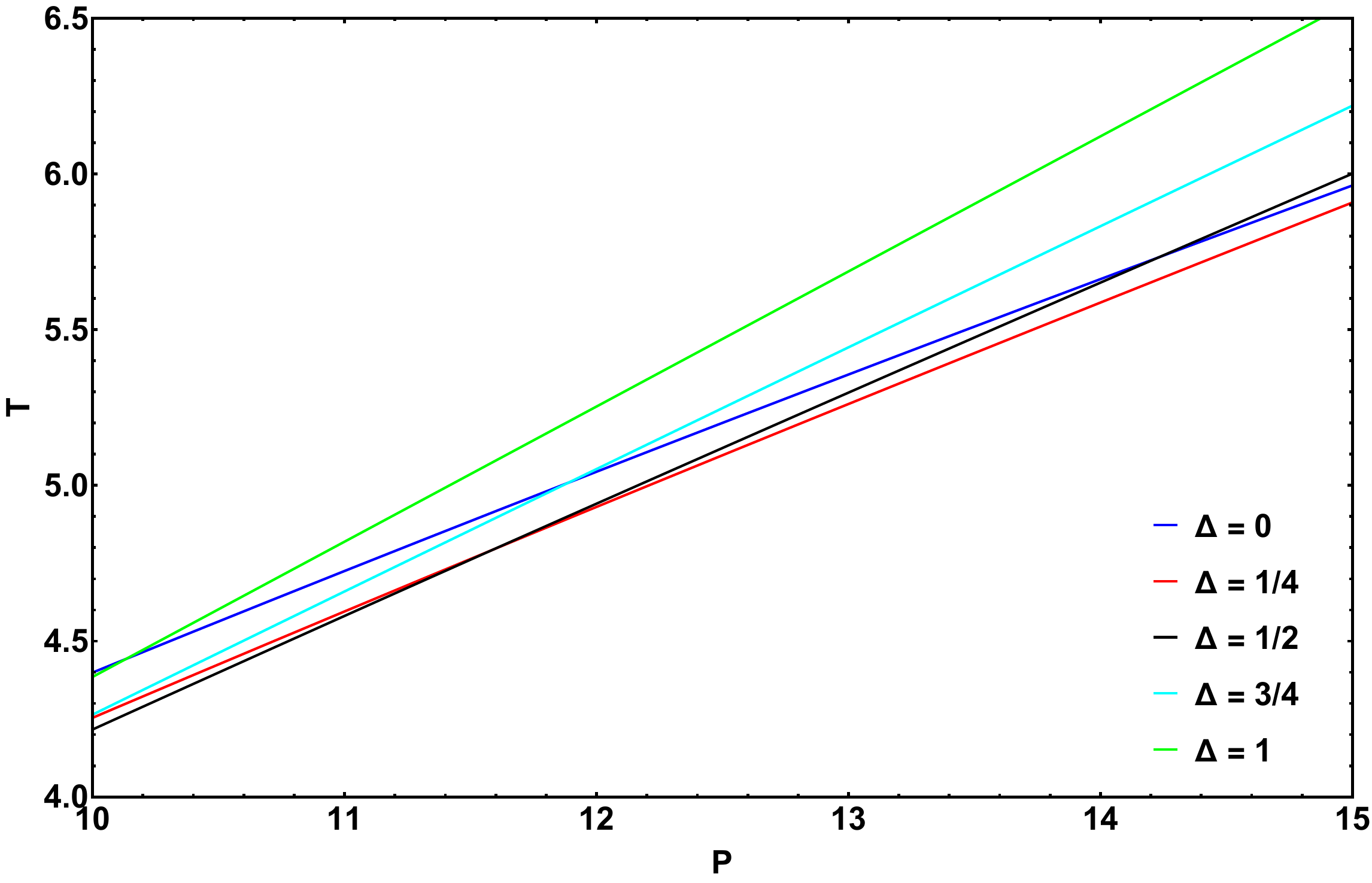}
\caption{Inversion temperature as a function of the pressure for fixed  $\omega = -1.0$, $c = 0.1$, $a = 0.1$, $Q = 1.0$, and the corrections to entropy: $\Delta = 0$ (blue line), $\Delta = 1/4$ (red line), $\Delta = 1/2$ (black line), $\Delta= 3/4$ (cyan line) and $\Delta=1$ (green line). The upper panels show small and large behaviors of $T_{\rm i}^{\rm JT} \times P$, while the lower panels depict different ranges of these functions, with the purpose of showing the intercept points of these curves for different values of $\Delta$.} 
\label{figure1}
\end{figure}

From now on, we will discuss what happens when we compute the variation of $\Delta$ in the face of the other parameters. First,  we will consider the case of fixed parameters $a=0.1, \, c=0.1,\, \omega=-1.0,\,  Q=1.0$ and we will vary $\Delta$ with a difference $1/4$ in the interval $\Delta\in[0,1]$. Fig.~\ref{figure1} shows the results of this analysis. In the upper panels, we plot the inversion temperature, Eq.  \eqref{inversiontemp},  against the pressure $P$ in small and large ranges. In these panels, one sees that increasing $\Delta$ reduces the inversion temperature $T_{\rm i}^{\rm JT}$, Eq. \eqref{inversiontemp}, for fixed $P$, or equivalently, increases $P$ for fixed $T_{\rm i}^{\rm JT}$. This comes with a change of slope in the curves, without altering the zeros of these functions. Actually, the five curves corresponding to the values of $\Delta$ in the interval $[0,1]$, have the same zero, which is very close to $P_0\approx0.001$. The results of the left panel of Fig.~\ref{figure1} about the decrease of the slope of the inversion temperature curves with increasing values of $\Delta$ are compatible with those found in Fig.~16 of \cite{Yasir:2024oah} with $F(R)$ gravity, with Figs.~3 and 4 of \cite{Javed:2025dit} within Rastall gravity, and with Fig.~4 of \cite{Ladghami:2024yjn} within spacetime foam effects.

Then, in the lower panels of Fig. \ref{figure1}, we show the behavior of these functions in the ranges $7\le P \le 12$ and $10\le P \le 15$, to emphasize the crossing of the inversion temperature curves corresponding to different $\Delta$ values. This means that increasing the fractal dimension $\Delta$ implies decreasing the inversion temperature $T_{\rm i}^{\rm JT}$ for pressures lower than a certain critical value $P_c$, and the opposite behavior above this value. In particular, in the lower left panel of Fig. \ref{figure1}, it is seen that the curves for $\Delta=0$ and  $\Delta=1$ cross at $P_c\approx 10.2$, which can be determined numerically from Eq.  \eqref{inversiontemp} using the values $a=0.1,\, c=0.1, \,\omega=-1.0,\, Q=1.0$, for $\Delta=0$ and  $\Delta=1$.

\begin{figure}[h]
\vskip0.5cm 
\centering
\includegraphics[width=7.8cm]{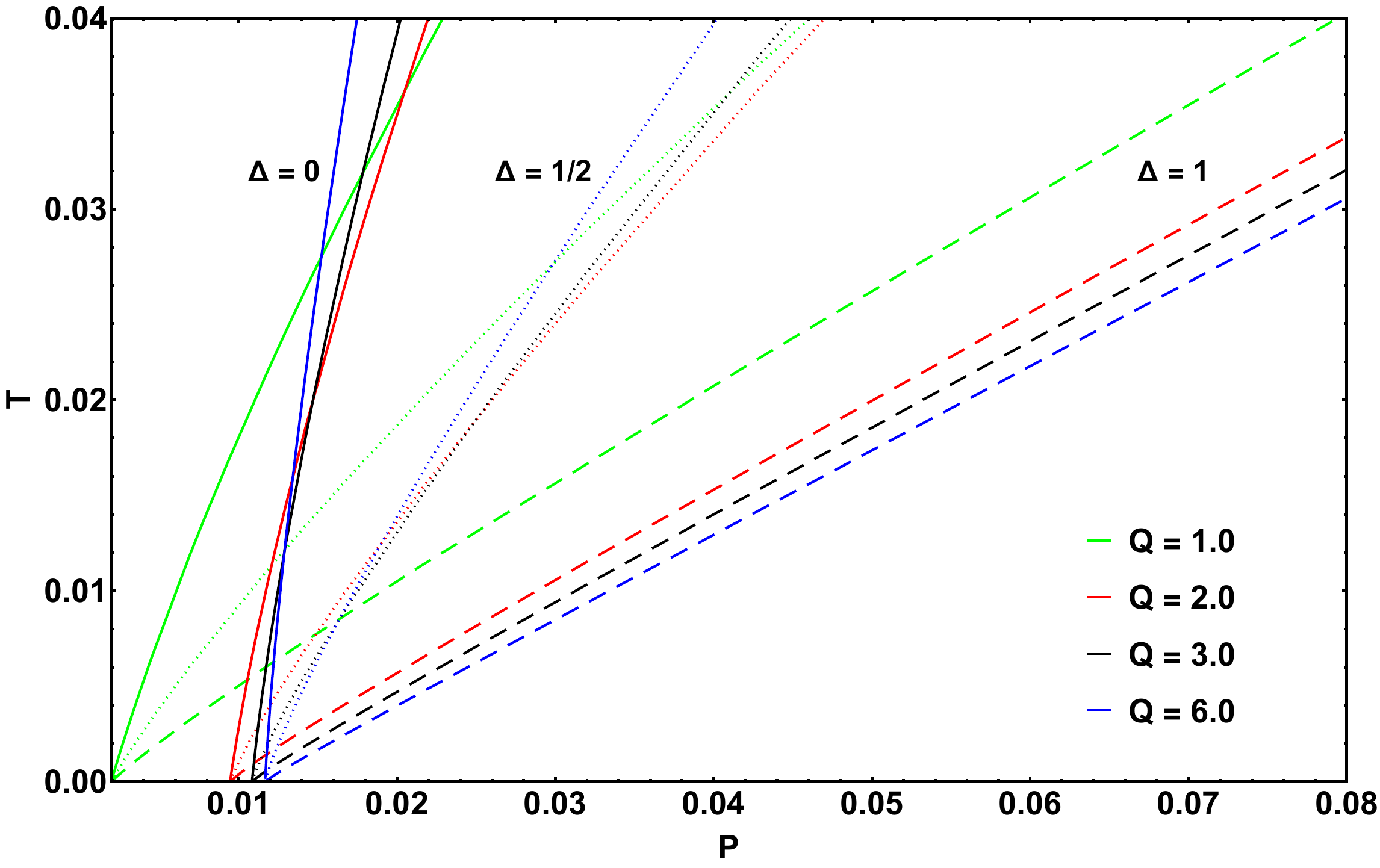}\qquad \includegraphics[width=7.6cm]{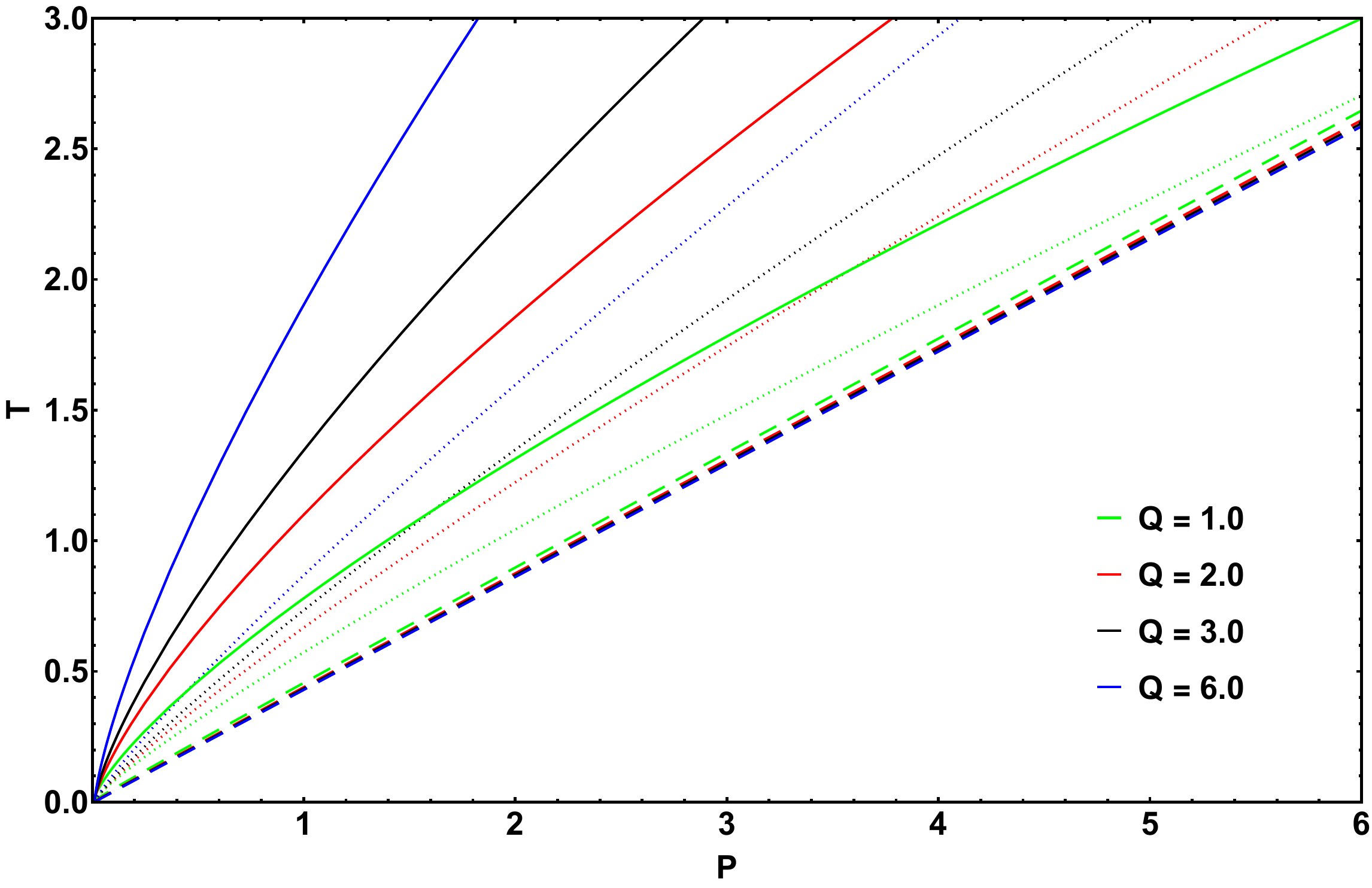}
\caption{Inversion temperature as a function of the pressure for fixed $\omega = -1.0$, $c = 0.1$, $a = 0.1$, small range, and the values for charge: $Q = 1.0$ (green line), $Q = 2.0$ (red line), $Q = 3.0$ (black line) and $Q= 6.0$ (blue line). Solid lines are obtained from $\Delta=0$, dotted lines from $\Delta=1/2$ and dashed lines from $\Delta=1$. \textit{Left panel}: small range. \textit{Right panel}: large range.}
\label{figure2}
\end{figure}

In Fig. \ref{figure2}, we show the results for $T_{\rm i}^{\rm JT}\times P$ on both small and large scales, using $\Delta = 0$, $\Delta = 1/2$, and  $\Delta = 1$, keeping $a=0.1, \, c=0.1,\, \omega=-1.0$ fixed and varying the charge $Q$ from 1 to 6. Note that increasing $\Delta$ diminishes the slope of the curves $T_{\rm i}^{\rm JT}\times P$, as already found in Fig. \ref{figure1}. In this picture, we also see that the effect on $T_{\rm i}^{\rm JT}$ of increasing the charge $Q$ is similar to the decrease in the fractal parameter of Barrow $\Delta$, so that from the thermodynamical point of view, the charge $Q$ and $\Delta$ have opposite (inverse) behaviors. Note also that the zeros of these curves are different for each charge value $Q$, corresponding to different initial pressures, and we see that these zeros are not modified by changing $\Delta$. For the small range (left panel), one clearly sees a distinction from the curves with $\Delta = 0$, $1/2$, and  $1$, while in the large range (right panel) these curves get mixed, since for $\Delta = 1$, the slope is almost constant, whilst for $\Delta = 0$, and $1/2$ they vary significantly, with decreasing slope with increasing pressure. The results of the right panel of Fig.~\ref{figure2} can be compared  with those found in Figs.~3 and 4 of \cite{Javed:2025dit} within Rastall gravity.

\begin{figure}[h]
\vskip0.5cm 
\centering
\includegraphics[width=7.8cm]{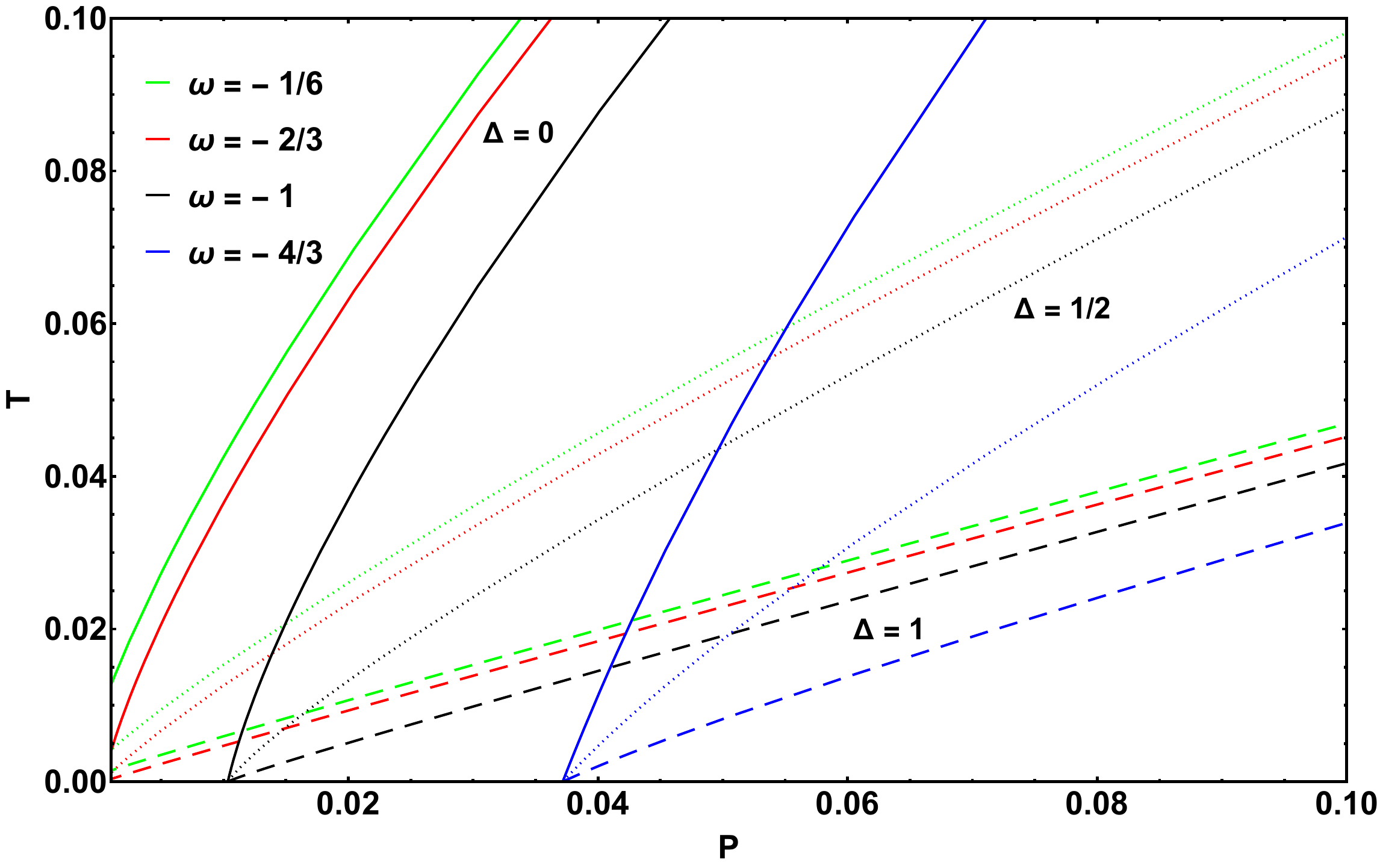}\qquad \includegraphics[width=7.8cm]{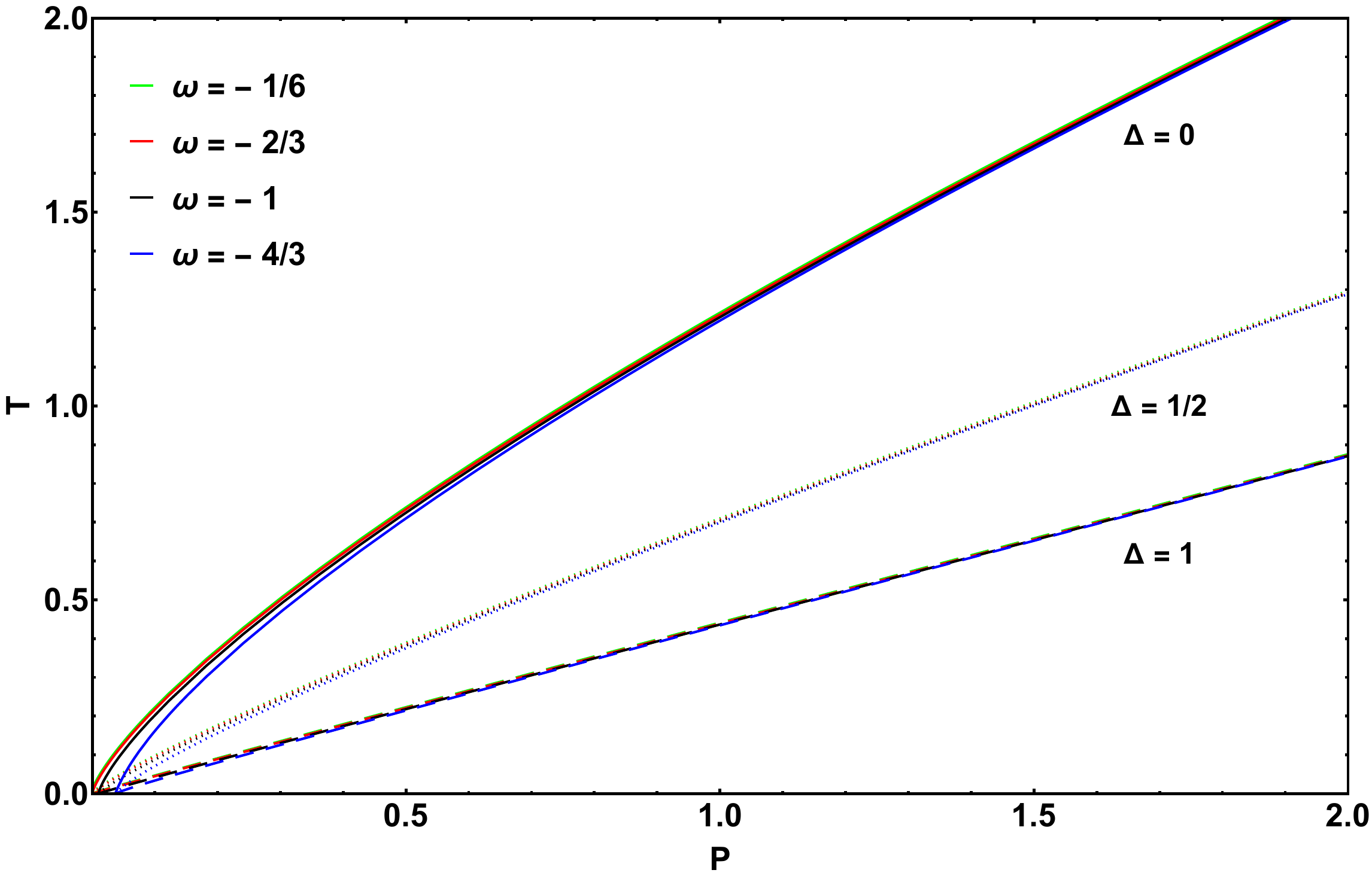}
\caption{Inversion temperature as a function of the pressure for fixed $Q = 2.5$, $c = 0.1$, $a = 0.1$, varying $\omega$: $\omega = -1.6$ (green line), $\omega = -2/3$ (red line), $\omega = -1$ (black line) and $\omega = -4/3$ (blue line). Solid lines are obtained from $\Delta=0$, dotted lines from $\Delta=1/2$ and dashed lines from $\Delta=1$. \textit{Left panel}: small range. \textit{Right panel}: large range.}
\label{figure3}
\end{figure}

In Fig. \ref{figure3}, we plot $T_{\rm i}^{\rm JT}\times P$ for fixed $Q = 2.5$, $c = 0.1$, $a = 0.1$, varying the Kiselev parameter  $\omega = -1.6$, $-2/3$, $-1$, and $-4/3$, for $\Delta=0$, 1/2 and 1. For fixed $\Delta$, the slope of the curves for these values of $\omega$ is very similar, although the zeros of the curves are clearly distinguishable. The slopes of these curves are very sensitive to the variation of $\Delta$, while the zeros are not. Again, increasing $\Delta$ decreases the slope of the curves $T_{\rm i}^{\rm JT}\times P$. In both the small range (left panel) and the large range (right panel), one easily sees a distinction between the curves with different values of $\Delta$, especially in the large range, where the curves with different values of $\omega$ seem to collapse mainly depending on $\Delta$.

\begin{figure}[h]
\vskip0.5cm 
\centering
\includegraphics[width=7.8cm]{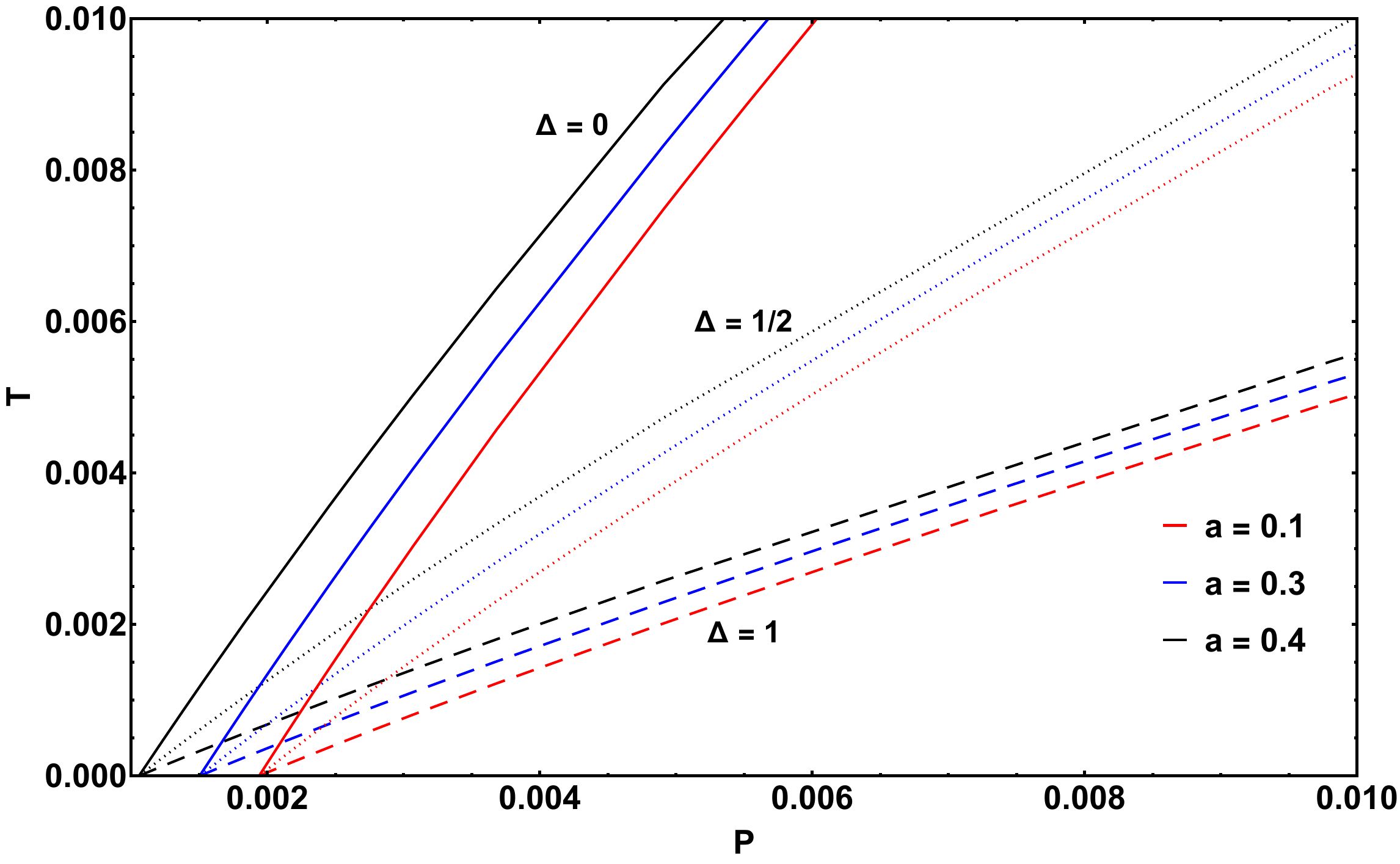}
\caption{Inversion temperature as a function of the pressure for fixed $\omega = -1.0$, $c = 0.1$, $Q = 1.0$,  and the values for the quantum corrections: $a = 0.1$ (blue line), $a = 0.3$ (red line), $a = 0.4$ (black line). Solid lines are obtained with $\Delta=0$, dotted with $\Delta=1/2$, and dashed lines with $\Delta=1$.}
\label{figure4}
\end{figure}

To study the effect of the variation of $\Delta$ and the quantum correction parameter $a$, we present in Fig. \ref{figure4} the graph $T_{\rm i}^{\rm JT}\times P$ for fixed $\omega = -1.0$, $c = 0.1$, $Q = 1.0$, varying $a = 0.1$, $0.3$ and $0.4$ while $\Delta=0$, 1/2 and 1. In this picture, one sees that the slopes of the curves decrease with increasing $\Delta$, whereas the corresponding zeros of $T_{\rm i}^{\rm JT}$ are not modified by the change of $\Delta$. These curves have approximately the same slopes independent of the values of $a$. On the other hand, increasing $a$ shifts the zeros of $T_{\rm i}^{\rm JT}$ to smaller values. For example, if $a=0.4$ the zero of $T_{\rm i}^{\rm JT}$ occurs at $P\approx 0.002$, and for $a=0.1$ it appears at $P\approx0.001$. In particular, these curves appear in distinct groups depending on the $\Delta$ values, at least for the range chosen here. 

\begin{figure}[h]
\vskip0.5cm 
\centering
\includegraphics[width=7.8cm]{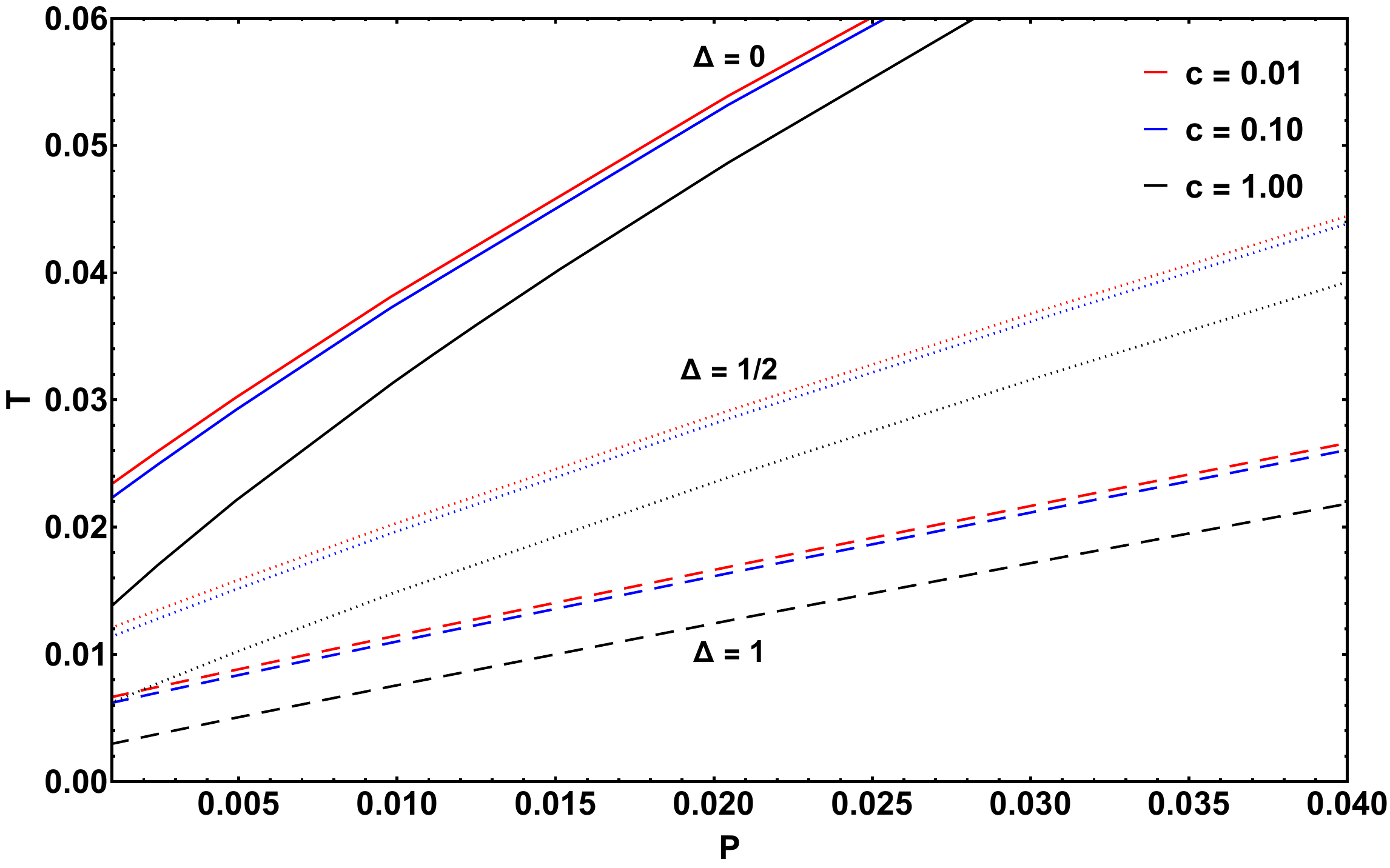}\qquad\includegraphics[width=7.8cm]{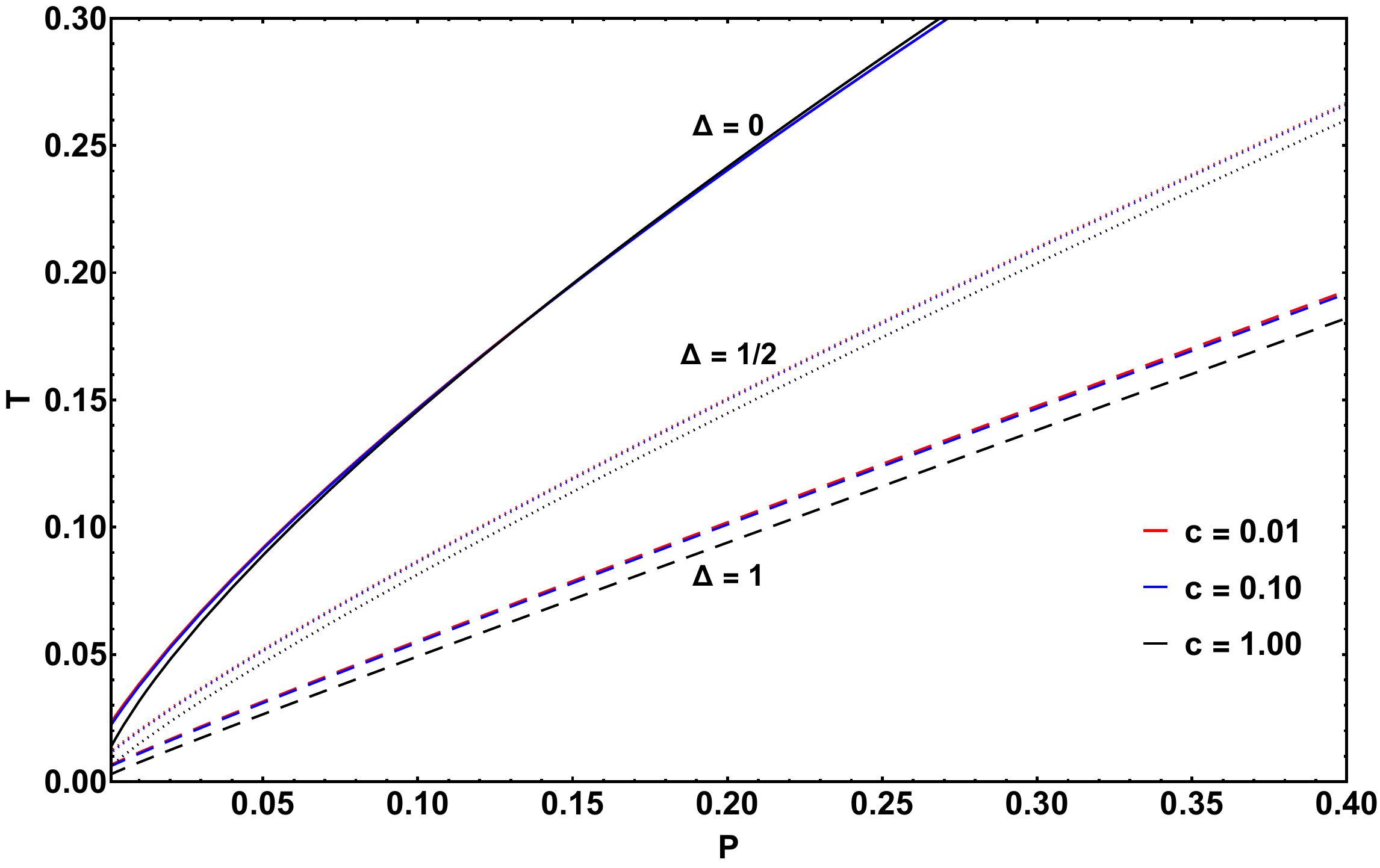}
\caption{Inversion temperature as a function of the pressure for fixed $\omega = -1/6$, $a = 0.1$, $Q = 1.0$, varying $c$: $c = 0.01$ (red line), $c = 0.10$ (blue line), $c = 1.00$ (black line). Solid lines are obtained with $\Delta=0$, dotted lines from $\Delta=1/2$ and dashed lines from $\Delta=1$.  \textit{Left panel}: small range. \textit{Right panel}: large range.}
\label{figure5}
\end{figure}

\begin{figure}[h]
\vskip0.5cm 
\centering
\includegraphics[width=7.8cm]{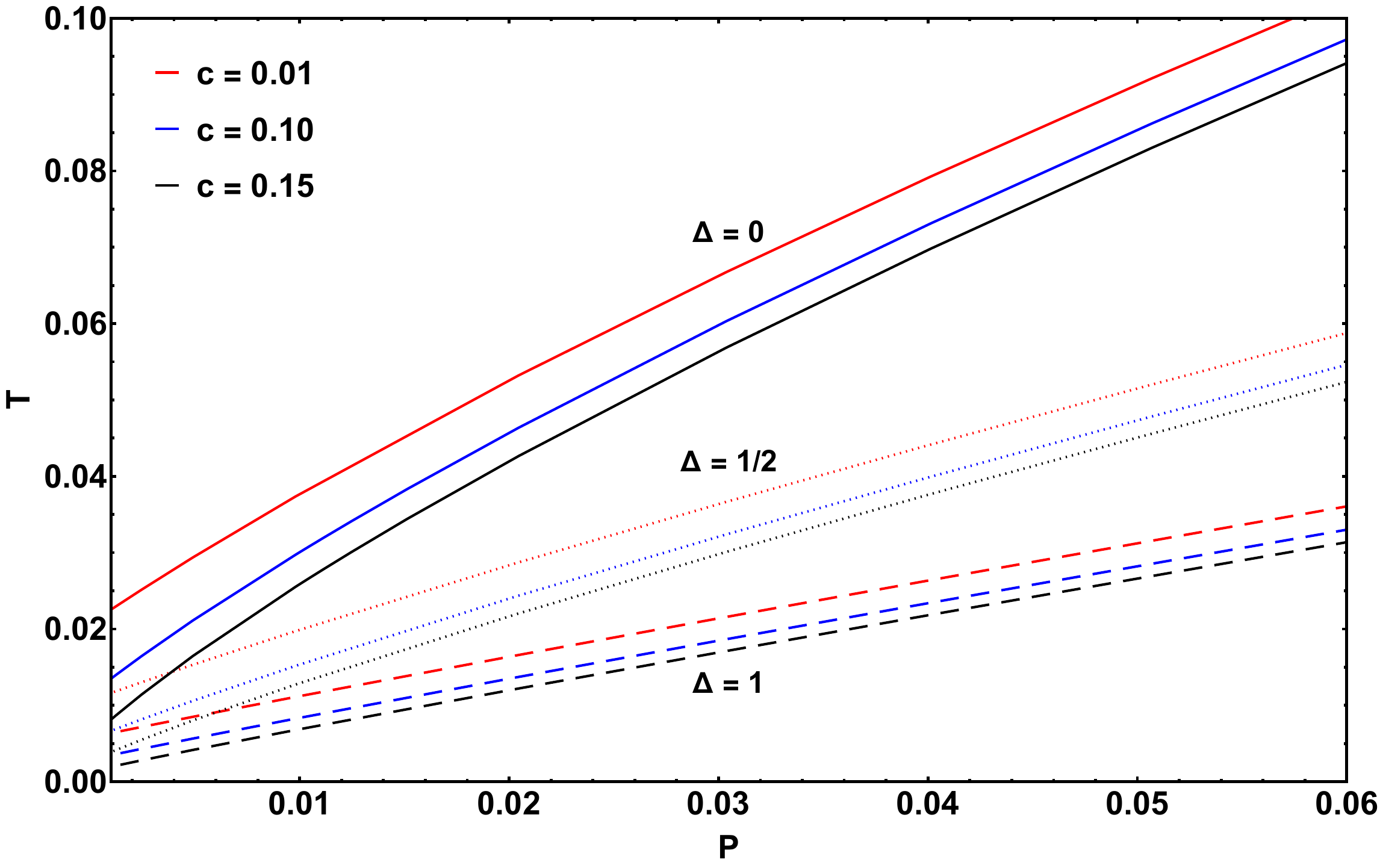}\qquad\includegraphics[width=7.66cm]{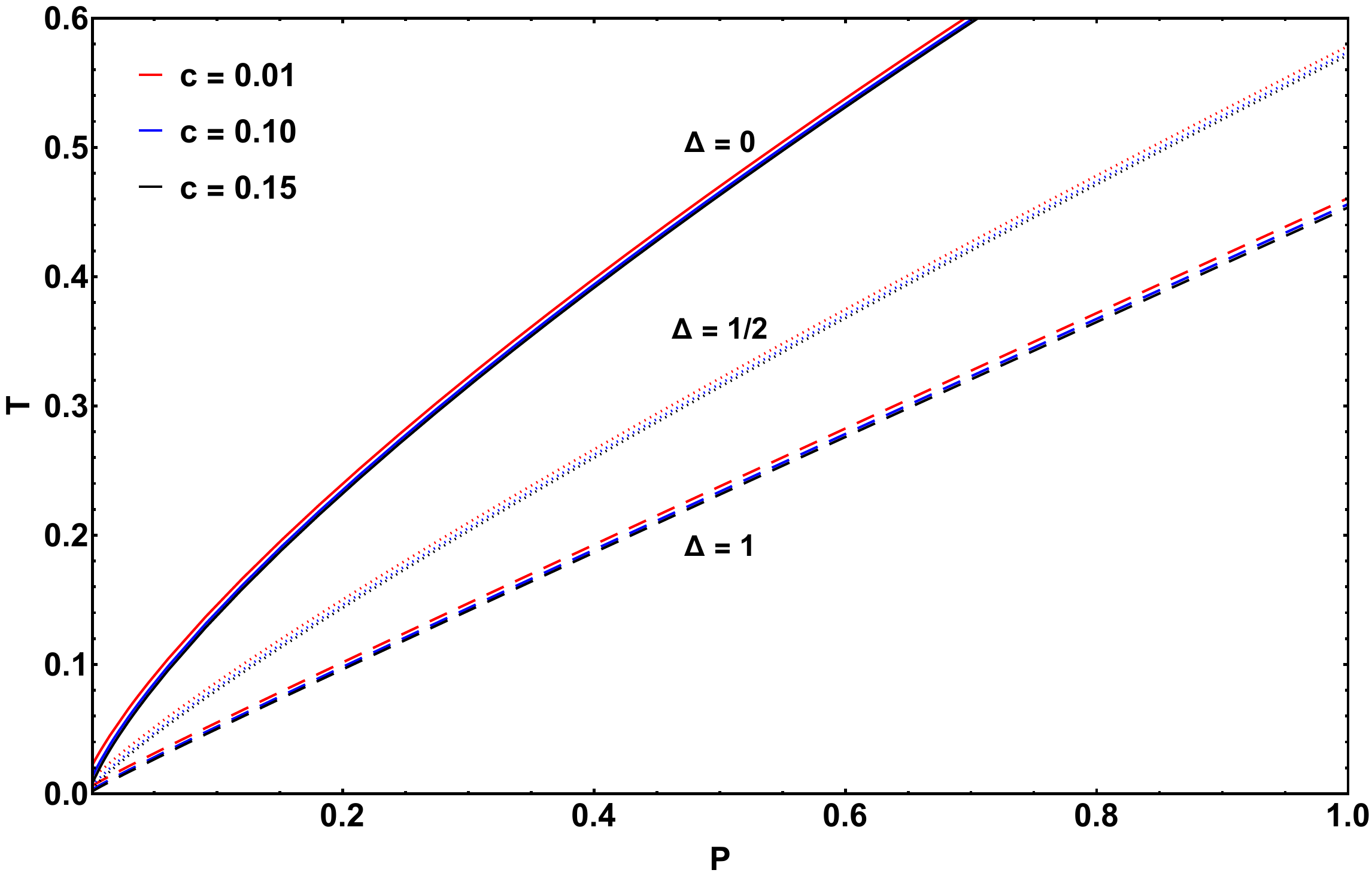}
\caption{Inversion temperature as a function of the pressure for fixed $\omega = -2/3$, $a = 0.1$, $Q = 1.0$, varying $c$: $c = 0.01$ (red line), $c = 0.10$ (blue line), $c = 0.15$ (black line). Solid lines are obtained with $\Delta=0$, dotted lines from $\Delta=1/2$ and dashed lines from $\Delta=1$.  \textit{Left panel}: small range. \textit{Right panel}: large range.}
\label{figure6}
\end{figure}

\begin{figure}[h]
\vskip0.5cm 
\centering
\includegraphics[width=7.8cm]{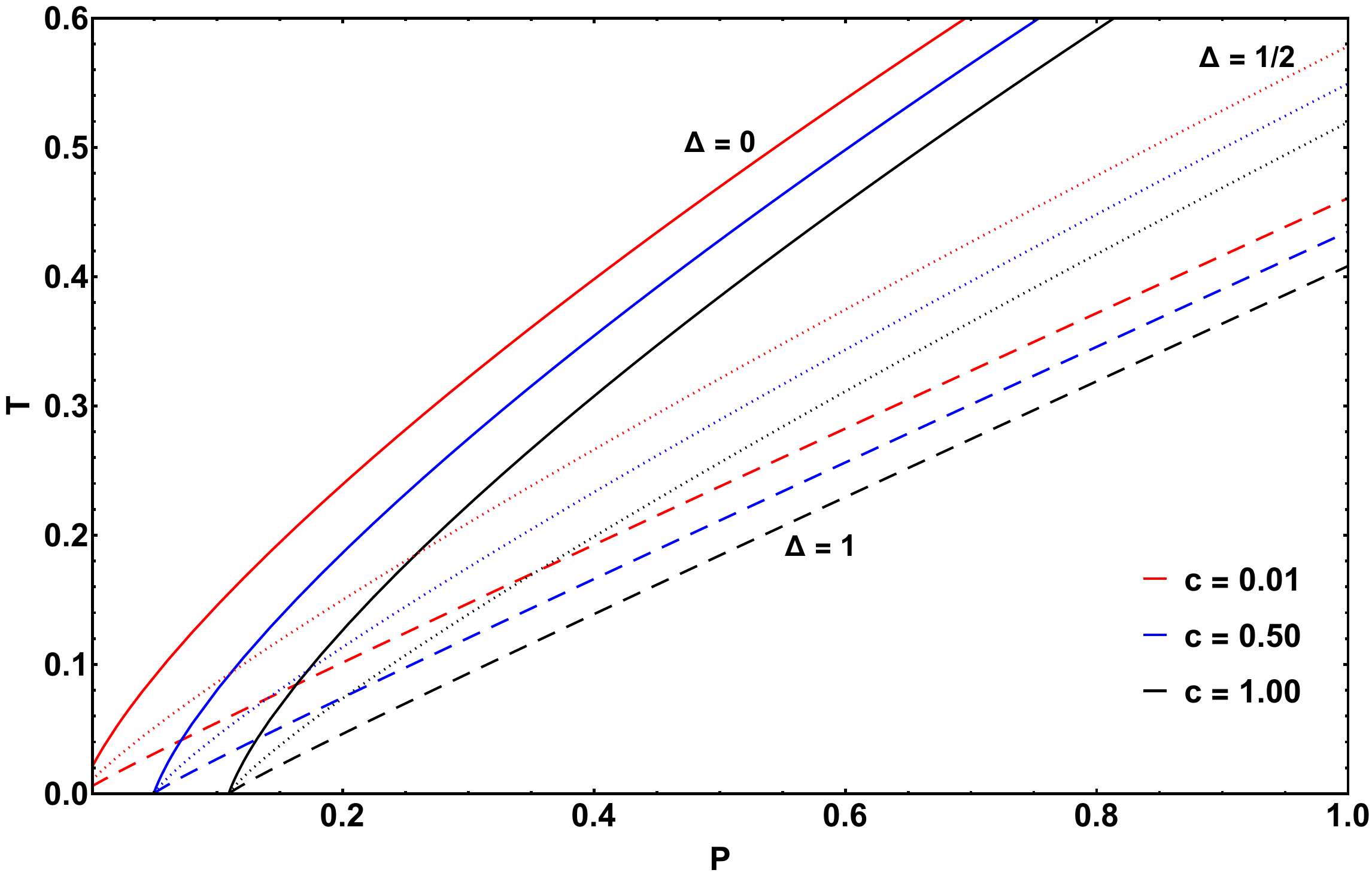}
\includegraphics[width=7.8cm]{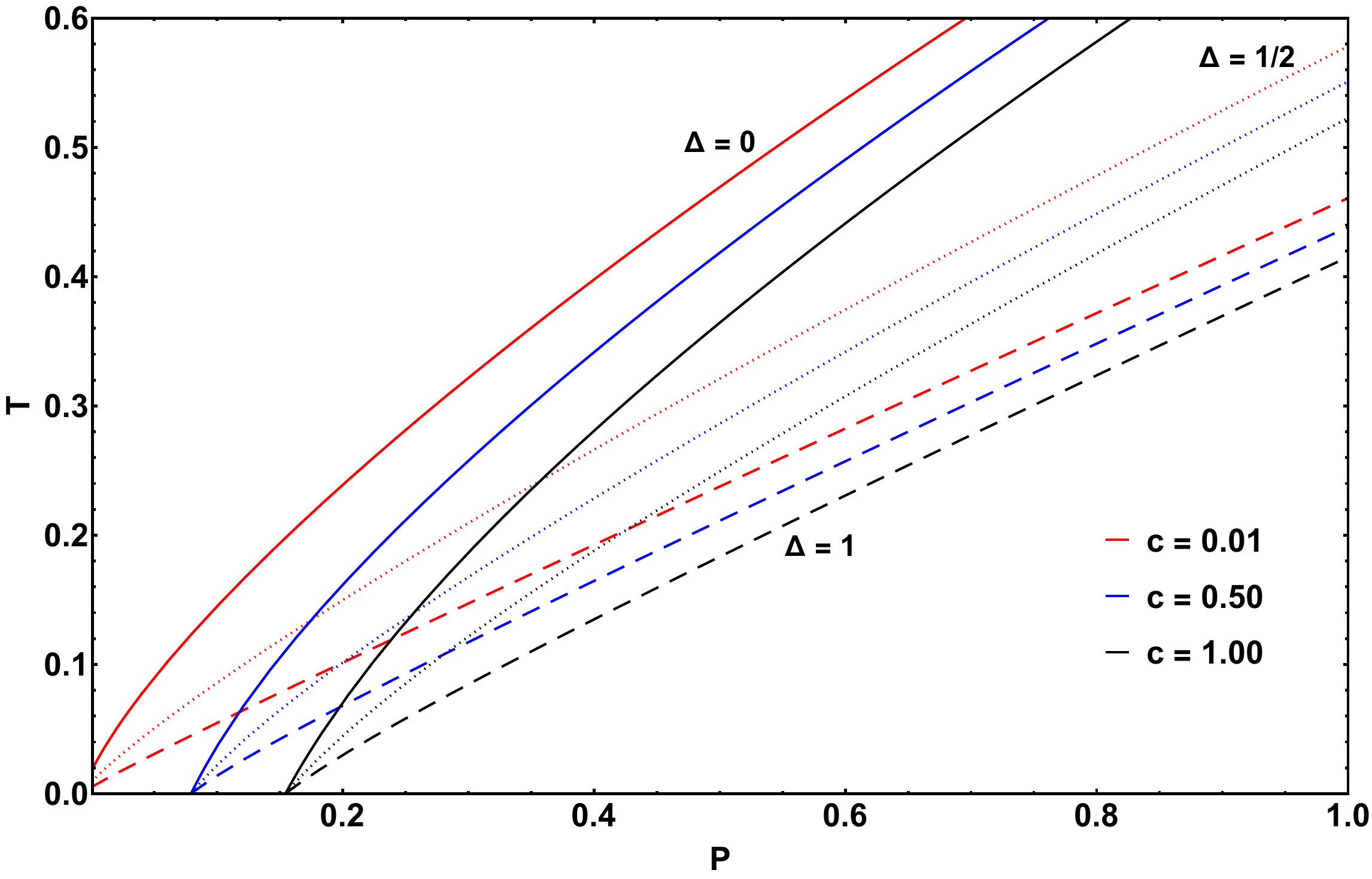}
\caption{Inversion temperature as a function of the pressure for fixed $\omega = -1$ (left panel) and $\omega = -4/3$ (right panel), both with $a = 0.1$, $Q = 1.0$, varying $c$: $c = 0.01$ (red line), $c = 0.50$ (blue line), $c = 1.00$ (black line). Solid lines are obtained with $\Delta=0$, dotted lines from $\Delta=1/2$ and dashed lines from $\Delta=1$.}
\label{figure7}
\end{figure}

Variations of the Kiselev coupling $c$ are presented in Figs. \ref{figure5}, \ref{figure6}, and \ref{figure7} for the curves $T_{\rm i}^{\rm JT}\times P$ with fixed $\omega = -1/6$ (Fig. \ref{figure5}), $\omega = -2/3$ (Fig. \ref{figure6}) and $\omega = -1$ (Fig. \ref{figure7} left panel), $\omega = -4/3$ (Fig. \ref{figure7} right panel), all of which with $a = 0.1$, and $Q = 1.0$, for $\Delta=0$, 1/2 and 1. In Fig. \ref{figure5}, the values of $c$ are 0.01, 0.10, and 1.00, while in Fig. \ref{figure6}, the values of $c$ are 0.01, 0.10 and 0.15, and in Fig. \ref{figure7}, the values of $c$ are 0.01, 0.50, and 1.00. In these three pictures, one can see that the increase in the values of $\Delta$ decreases the slope of the inversion temperature curves. The values $\Delta=0$, 1/2 and 1 are apart enough to clearly see the distinction between these groups of curves. In Figs. \ref{figure5} and \ref{figure6}, there are minimum values for $T_{\rm i}^{\rm JT}$ that correspond to $P=0$. This is a consequence of the values $\omega = -1/6$ (Fig. \ref{figure5}), and $\omega = -2/3$ (Fig. \ref{figure6}). The increasing values of $c$ and $\Delta$ both imply a decrease in $T_{\rm i}^{\rm JT}$ for $P=0$. In Fig. \ref{figure7}, the lowest value of $c=0.01$ still implies a finite $T_{\rm i}^{\rm JT}$ for $P=0$ in both panels, but for $c=0.50$ and $c=1.00$ the minimum value for $T_{\rm i}^{\rm JT}$ is zero corresponding to non-zero pressures. In this case, the increasing value of $c$ increases the pressure at zero $T_{\rm i}^{\rm JT}$. In the three pictures, \ref{figure5}, \ref{figure6}, and \ref{figure7}, the curves for fixed $\Delta$ have almost the same slopes and are easily distinguishable from other values of $\Delta$, at least when the range is small. For larger ranges, curves with different $c$ but the same $\Delta$ tend to collapse.

\begin{figure}[h]
\vskip0.5cm 
\centering
\includegraphics[width=7.6cm]{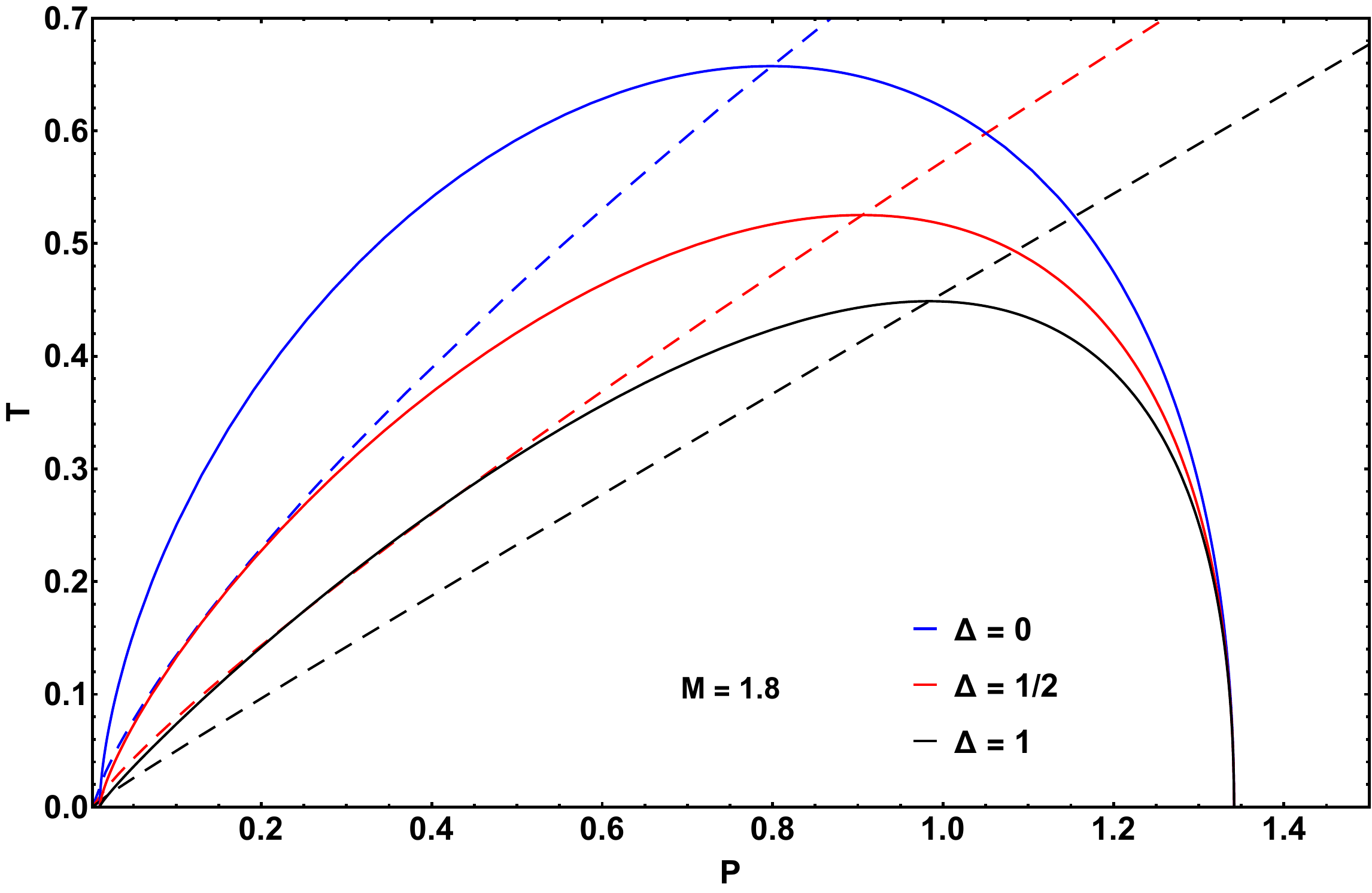}
\includegraphics[width=7.8cm]{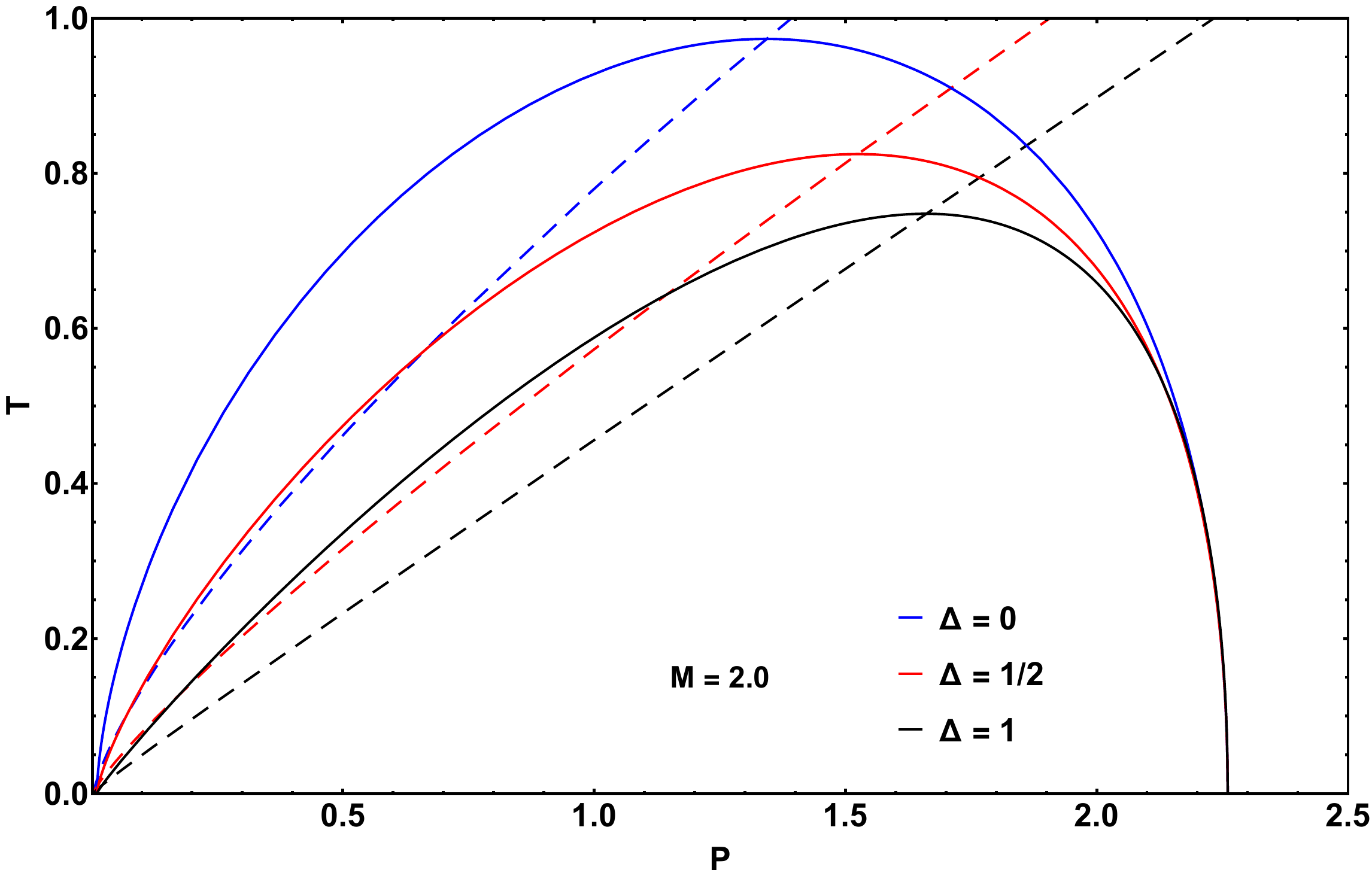}
\includegraphics[width=7.8cm]{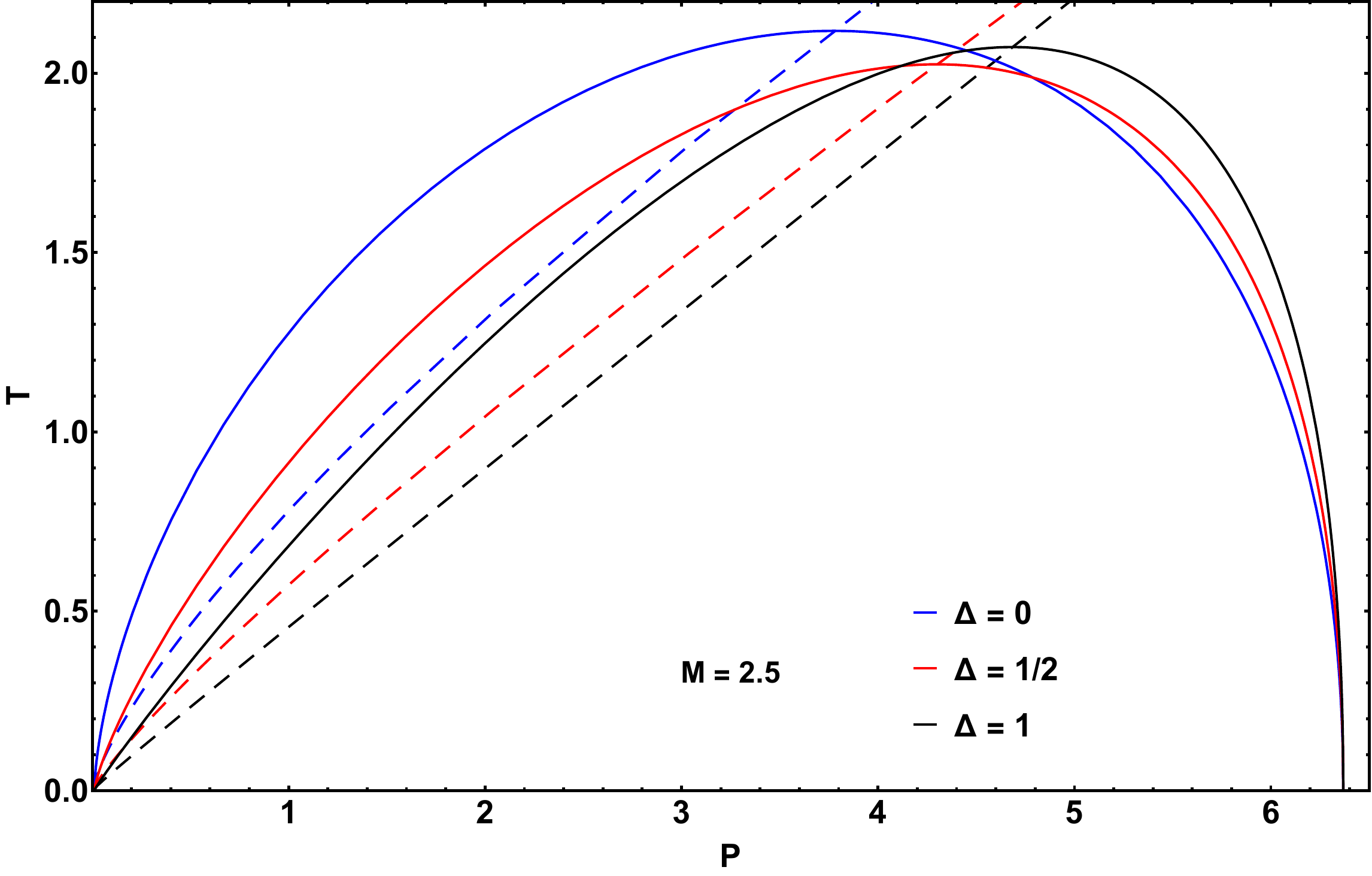}
\includegraphics[width=7.6cm]{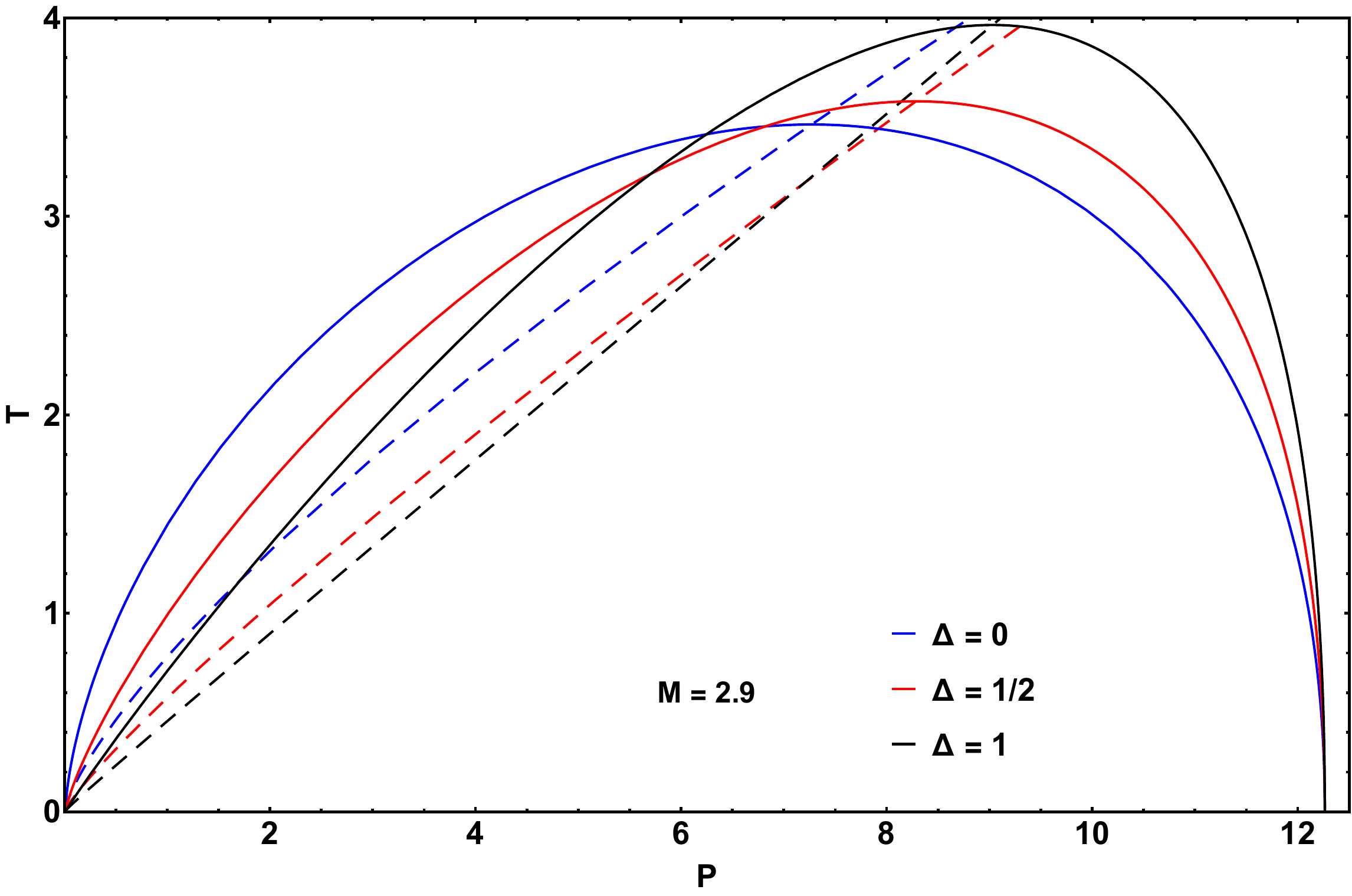}
\caption{This picture shows the isenthalpic curves as a function of the pressure as solid lines with fixed values of $M=1.8$ (upper left panel), $M=2.0$ (upper right panel), $M=2.5$ (lower left panel), and $M=2.9$ (lower right panel), with fixed $\omega = -1$, $a = 0.1$, $Q = 1.0$, and $c = 0.1$, for different values of the correction in entropy: $\Delta=0$ (blue line), $\Delta = 1/2$ (red line), $\Delta = 1$ (black line). Note that these curves have common zeros independent of the $\Delta$ values. Inversion temperature curves from Eq. (\ref{inversiontemp}) are represented by dashed lines for the same choice of parameters. These panels also show a transition behavior around $M=2.5$. For lower masses ($M<2.5$, upper panels) the increase of $\Delta$ decreases the isenthalpic curves, and for $M\ge 2.5$ the increase of $\Delta$ increase the isenthalpic curves.}
\label{figure8}
\end{figure}

\begin{figure}[h]
\vskip0.5cm 
\centering
\includegraphics[width=7.8cm]{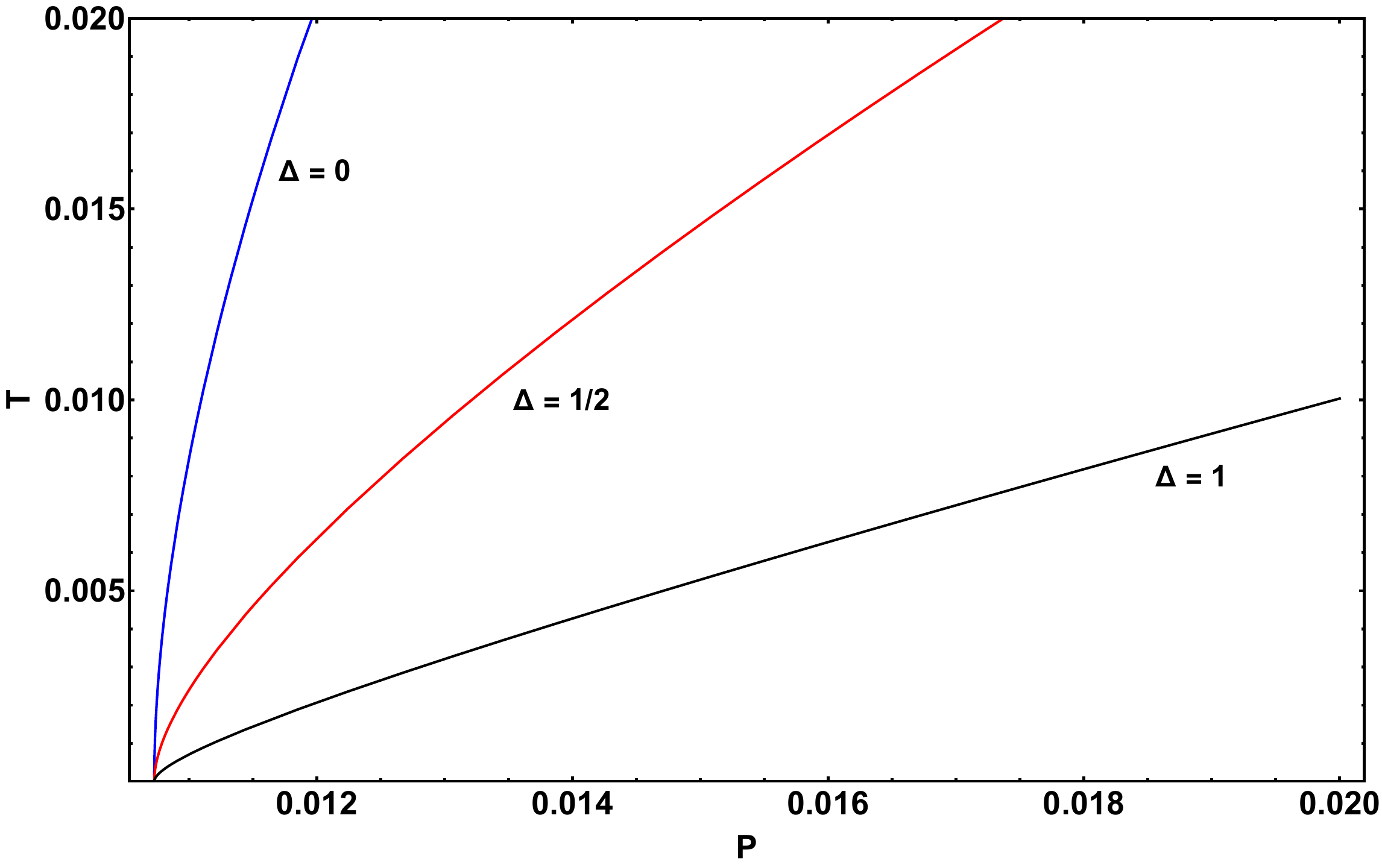}\qquad\includegraphics[width=7.8cm]{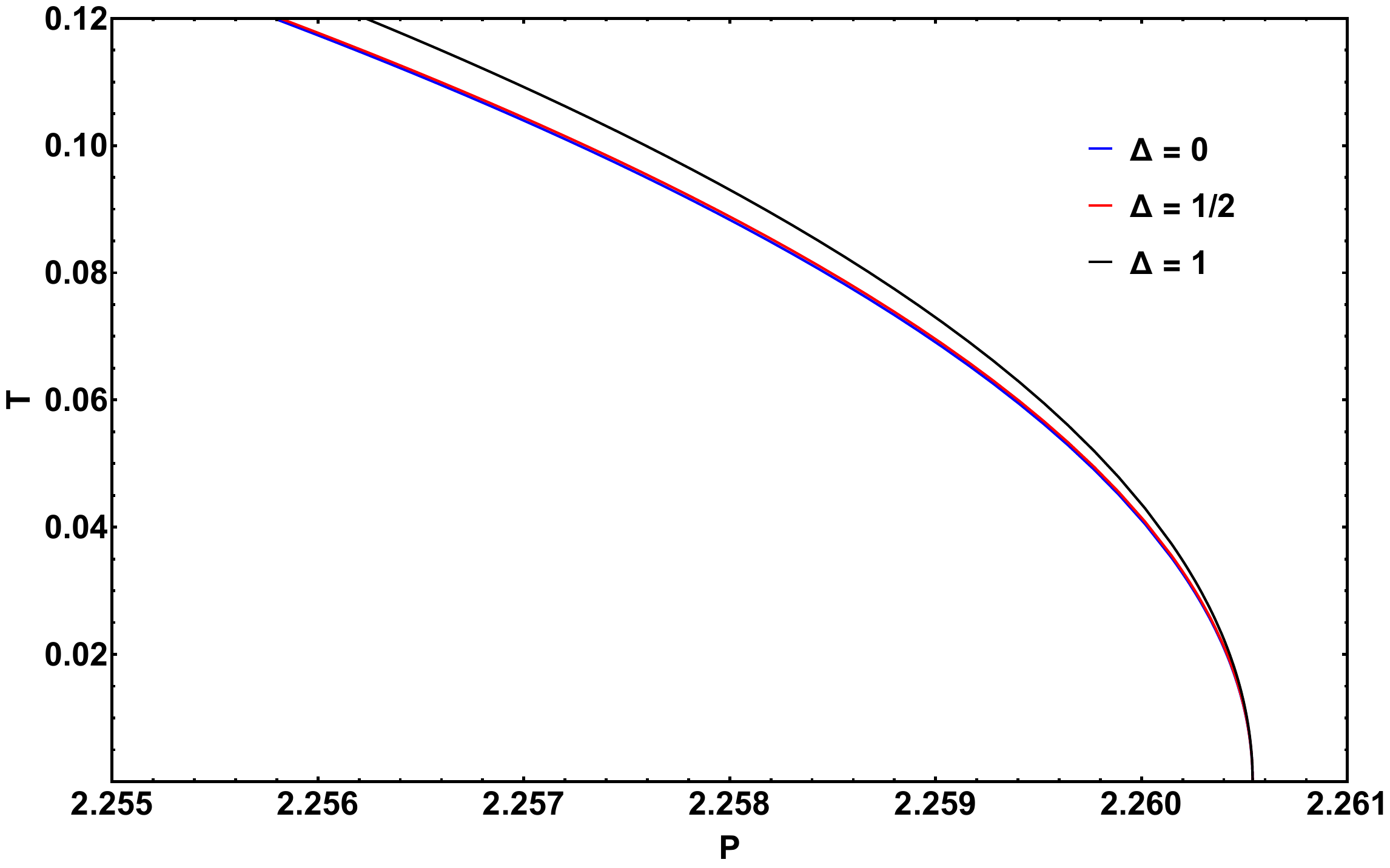}
\caption{Isenthalpic curves, obtained with the parameters $\omega = -1$, $a = 0.1$, $Q = 1.0$, $c = 0.1$, $M=2.0$. Here, we show that the two zeros in the $T\times P$ plane for the isenthalpics are identical for a given fixed mass $M$ and independent, as expected, of $\Delta$.}
\label{figure9}
\end{figure}

\begin{figure}[h]
\vskip0.5cm 
\centering
\includegraphics[width=7.8cm]{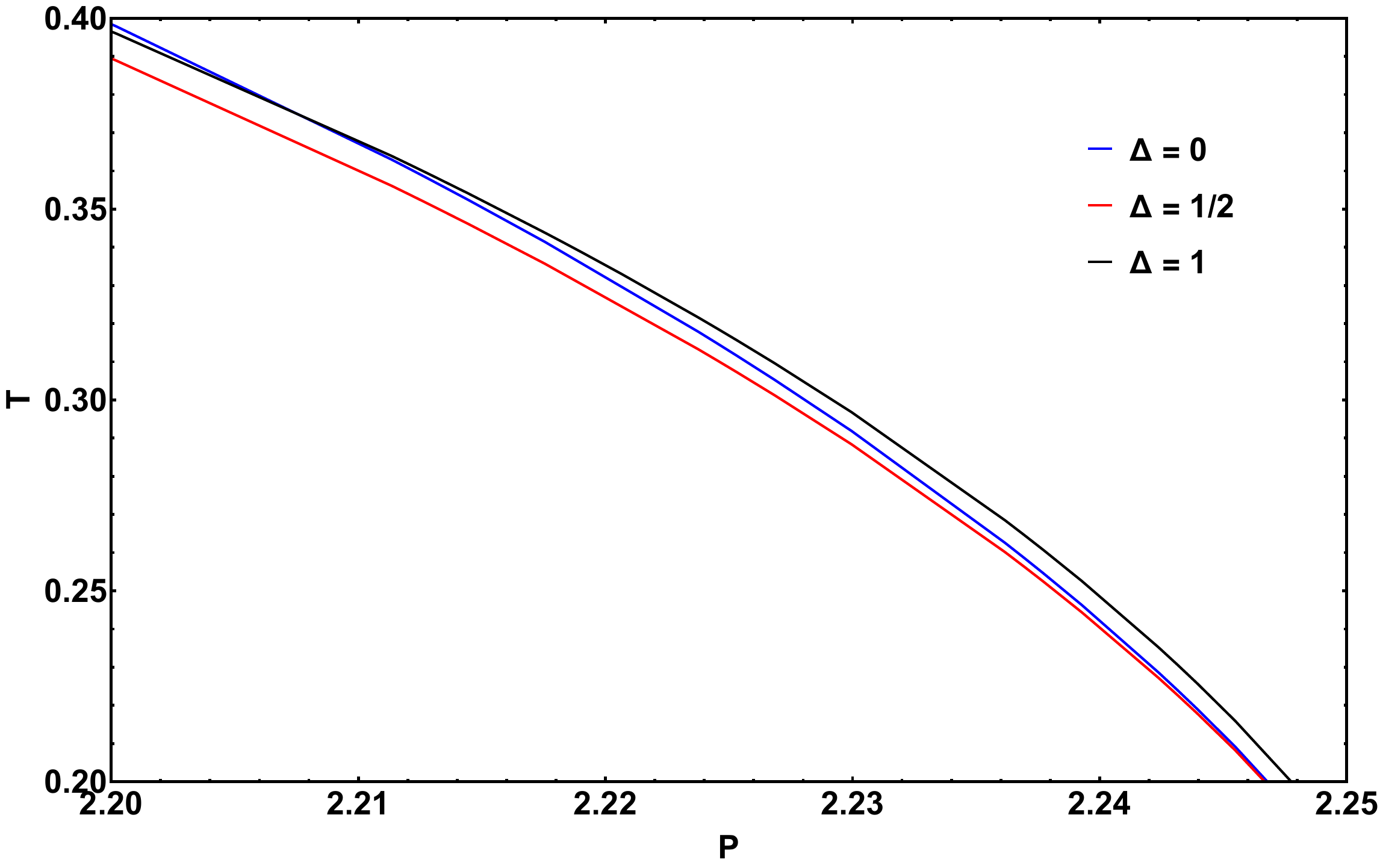}
\caption{Isenthalpic curves, obtained with the parameters $\omega = -1$, $a = 0.1$, $Q = 1.0$, $c = 0.1$, $M=2.0$. Here, we show that the curves intersect at different values between the zeros.}
\label{figure10}
\end{figure}

 The isenthalpic curves can be constructed by fixing a mass $M$, as given by Eq. \eqref{Mass}, rewritten in terms of pressure $P$, as defined by Eq. \eqref{Pl}. Then, pressure $P$ is found as a function of the external radius of BH, $r_+$ and the parameters $\omega$, $a$, $Q$ and $c$. This relation is, in general, transcendental and requires numerical treatment, so once the parameters are fixed, a table of values $P\times r_+$ can be constructed. Substituting the values of $r_+$ in Eq. \eqref{temp}, one finds a function of the temperature $T$ in terms of the pressure $P$, for fixed mass, which are the isenthalpics. Since Eq. \eqref{temp} also depends on $\Delta$, so do the isenthalpic curves. In Fig. \ref{figure8}, we present these  curves for $M=1.8$, $M=2.0$, $M=2.5$, and $M=2.9$ while we fixed $\omega = -1$, $a = 0.1$, $Q = 1.0$ and $c = 0.1$, selecting values of  $\Delta=0$, 1/2, and 1.  
 Note that the results for the upper panels ($M=1.8$, $M=2.0$), the increase of $\Delta$ decreases the corresponding isenthalpic curves. The situation is opposite for the lower panels ($M=2.5$, and $M=2.9$) where increasing $\Delta$ also increases the isenthalpics. 
 The results of the lower panels of Fig.~\ref{figure8} can be compared with Figs.~5-10 of \cite{Javed:2025dit}, within Rastall gravity.

 Note that the isenthalpic curves in Fig.~\ref{figure8}, both in the upper and lower panels, for different values of $\Delta$ have two common zeros for each mass of BH, as shown in detail in the two panels of Fig. \ref{figure9} for $M=2.0$. The values of these zeros can also be obtained from our analysis taking $T=0$ in Eq. \eqref{temp}, 
 \begin{equation}\label{temp=0}
     0=\frac{1}{2(2+\Delta)(\sqrt{\pi} \r+)^{\Delta}}\bigg(8 P\r+-\frac{Q^2}{\pi \r+^3}+\frac{1}{\pi\sqrt{\r+^2-a^2}}+\frac{3\omega c}{\pi \r+^{2+3\omega}}\bigg)
 \end{equation}
from which one sees that these zeros are independent of $\Delta$. Using the relation between the BH mass and the radius of the external horizon, Eq. \eqref{Mass}, with a given fixed mass, we find an equation for $r_+$, the roots are $r_{+1}=0.100001$,  $r_{+2}=0.352294$, $r_{+3}=5.64764$, for $M=2$, $\omega = -1$, $a = 0.1$, $Q = 1.0$, and $c = 0.1$. The first root $r_{+1}$ implies a negative pressure $P_1=-716$, so it is not physically realizable. The other two roots imply the pressures $P_2=2.26$, and $P_3=0.0107$, which correspond to the values seen in Figs. \ref{figure8} and \ref{figure9} for $T=T_{\rm i}^{\rm JT}=0$. 

The isenthalpic curves for different values of $\Delta$ also cross at different points in the graph $T\times P$. This is illustrated in Fig. \ref{figure10}, where one can see that the curves for $\Delta=0$ and $\Delta=1$ cross at $P\approx2.21$ and $T\approx0.37$. On the other hand, the curves with $\Delta=1/2$ and $\Delta=1$ merge in $P\approx 2.246$ and $T\approx 0.21$. These crossings can be obtained numerically from Eqs. \eqref{Mass}, \eqref{Pl}, and \eqref{temp} with the appropriate values of $\Delta$ and the other parameters, to get a function $T(P)$, and imposing that they have common values of $P$ and $T$ for different values of $\Delta$.


\FloatBarrier
\section{Conclusions and final considerations}
\label{Sec:Conc}

In this work, we have investigated the effect of the modification of Barrow on Bekenstein-Hawking black hole entropy, speccifically on the Joule-Thomson inversion temperature and isenthalpic curves in AdS-Reissner-Nordström black holes in Kiselev spacetime with quantum corrections. We saw that choosing different values of the Barrow fractal parameter $\Delta$ produces modifications in general, reducing the JT inversion temperature and increasing the corresponding pressure. 

After obtaining the general equations for the mass of the BH, Eq. \eqref{Mass}, its temperature, Eq. \eqref{temp}, and the Joule-Thomson inversion temperature, Eq. \eqref{inversiontemp}, we proceed to a numerical study of the problem, fixing some parameters and varying others. In particular, in Fig. \ref{figure1}, we considered the JT inversion temperature as a function of the pressure for fixed  $\omega$, $c$, $a$, $Q$, varying the Barrow fractal parameter $\Delta$: 0, 1/4, 1/2, 3/4 and 1. In this picture, it is clearly seen that the increase in the values of $\Delta$ decreases the inversion temperature and increases the pressure. This picture also shows that the $T_{\rm i}^{\rm JT}\times P$ curves cross for different values of $\Delta$. 

In Fig. \ref{figure2}, we examined the behavior of the curves $T_{\rm i}^{\rm JT}\times P$, fixing the parameters $\omega$, $c$, $a$, while varying the values of the BH charge $Q$: 1, 2, 3, and 6. Apart from a nontrivial characteristic of these curves at very low temperature and pressure (left panel), the general profile (right panel) is that increasing the values of $\Delta$ reduces $T_{\rm i}^{\rm JT}$ increasing the pressure. The effect of increasing the BH charge is the opposite: it increases $T_{\rm i}^{\rm JT}$ and decreases $P$. In addition, the increase of $\Delta$ tends to yield more straight curves than that of low values of $\Delta$. This is basically the same behavior found in Figs. \ref{figure3}-\ref{figure7}, where we vary different parameters and the values of $\Delta$. 

 We have also studied the isenthalpic curves $T\times P$ with $M$ fixed. As shown in Figs. \ref{figure8}-\ref{figure10}, these curves are clearly distinct for different values of $\Delta$, although they start and end at the same $T\times P$ points, as a consequence of the dependence on $\Delta$ in Eq. \eqref{temp}. In particular, in Fig. \ref{figure8} we find that for values of $M<2.5$ (upper panels) the increase of $\Delta$ decreases the isenthalpic curves, while for $M\ge2.5$ (lower panels) they increase the curves. 

 It is also interesting to mention that for the various Kiselev scenarios described by the examined values of $\omega$, the effect of the Barrow parameter $\Delta$ is essentially the same, as can be seen in Figs. \ref{figure3} and \ref{figure5}-\ref{figure7}. Despite some particular behavior for the small temperature and pressure regime, for large values of these quantities, as the values of $\Delta$ increase, the slope of the $T_{\rm i}^{\rm JT}$ curve decreases, implying an increase in pressure. This can be understood since $T=(\partial M/\partial S)_{P,Q}$, Eq. \eqref{T}, so that increasing the entropy $S$, should decrease $T$. 

There are various types of quantum corrections in gravity which depend on the specific situation or energy and length scales. In particular, in the case of Fig. \ref{figure4}, we clearly saw the numerical difference between the manifestations of the quantum corrections studied here, the one associated with parameter $a$ inside the metric, Eq. \eqref{f(r)}, and the other related to $\Delta$ appearing in the expression of entropy of the so-called Barrow entropic model, Eq. \eqref{S_Delta}. 
 The metric quantum correction deals with the geometry of the spacetime fabric and, on the other hand, the entropic quantum correction is related to the storage of the information in the event horizon which reflects a non smooth surface, a fractal one. The latter is connected to the quantum microstates of the BH. The variation of $\Delta$ changes the inclination of the curves and the variation of the $a$-parameter dislocated the initial points of each curve on the pressure scale. Since these corrections appear clearly in Fig. \ref{figure4}, one can see that both perturbations are consistent and with similar energy or length scales for the chosen values of the parameters. 

 As final remarks, it would be interesting to analyze the interplay of the BH microstates, for instance, from a string theory point of view, and the present effects studied on the BH thermodynamics, in particular in the Joule-Thomson inversion temperature.  Another final observation is that all the material discussed here disregards spins, but recently a fermionic JTE was reported \cite{Ji:2023epw}, so an extension to this case can also be considered.

\begin{acknowledgments} 
RACS is supported by Coordenação de Aperfeiçoamento de Pessoal de Nível Superior (CAPES). HBF is partially supported by Conselho Nacional de Desenvolvimento Cient\'{\i}fico e Tecnol\'{o}gico (CNPq) under grant  310346/2023-1, and Fundação Carlos Chagas Filho de Amparo à Pesquisa do Estado do Rio de Janeiro (FAPERJ) under grant E-26/204.095/2024.
\end{acknowledgments}



\begin{thebibliography}{99}

\section*{References}

\bibitem{Bekenstein:1973ur}
J.~D.~Bekenstein,
``Black holes and entropy,''
Phys. Rev. D \textbf{7}, 2333 (1973), 
doi:10.1103/PhysRevD.7.2333.

\bibitem{Hawking:1974sw}
S.~W.~Hawking,
``Particle Creation by Black Holes,''
Commun. Math. Phys. \textbf{43}, 199 (1975), 
[erratum: Commun. Math. Phys. \textbf{46} (1976), 206]
doi:10.1007/BF02345020.

\bibitem{Mann:2025xrb}
R.~B.~Mann,
``Black hole chemistry: The first 15 years,''
Int. J. Mod. Phys. D \textbf{34}, no.09, 2542001 (2025),
doi:10.1142/S0218271825420015
[arXiv:2508.01830 [gr-qc]].

\bibitem{Wald:2025nbz}
R.~M.~Wald,
``The Entropy of Black Holes,''
Gen. Rel. Grav. \textbf{57}, no.5, 87 (2025),
doi:10.1007/s10714-025-03417-x.

\bibitem{Elizalde:2025iku}
E.~Elizalde, S.~Nojiri and S.~D.~Odintsov,
``Black Hole Thermodynamics and Generalised Non-Extensive Entropy,''
Universe \textbf{11}, no.2, 60 (2025),
doi:10.3390/universe11020060
[arXiv:2502.05801 [gr-qc]].

\bibitem{Witten:2024upt}
E.~Witten,
``Introduction to black hole thermodynamics,''
Eur. Phys. J. Plus \textbf{140}, no.5, 430 (2025),
doi:10.1140/epjp/s13360-025-06288-y
[arXiv:2412.16795 [hep-th]].

\bibitem{Ruppeiner:2023wkq}
G.~Ruppeiner and A.~M.~Sturzu,
``Black hole microstructures in the extremal limit,''
Phys. Rev. D \textbf{108}, no.8, 086004 (2023),
doi:10.1103/PhysRevD.108.086004
[arXiv:2304.06187 [gr-qc]].

\bibitem{Ong:2022frf}
Y.~C.~Ong,
``On black hole thermodynamics, singularity, and gravitational entropy,''
Gen. Rel. Grav. \textbf{54}, no.10, 132 (2022),
doi:10.1007/s10714-022-03008-0
[arXiv:2210.16856 [gr-qc]].

\bibitem{Wei:2022dzw}
S.~W.~Wei, Y.~X.~Liu and R.~B.~Mann,
``Black Hole Solutions as Topological Thermodynamic Defects,''
Phys. Rev. Lett. \textbf{129}, no.19, 191101 (2022),
doi:10.1103/PhysRevLett.129.191101
[arXiv:2208.01932 [gr-qc]].

\bibitem{Auffinger:2022khh}
J.~Auffinger,
``Primordial black hole constraints with Hawking radiation{\textemdash}A review,''
Prog. Part. Nucl. Phys. \textbf{131}, 104040 (2023),
doi:10.1016/j.ppnp.2023.104040
[arXiv:2206.02672 [astro-ph.CO]].

\bibitem{Bravo-Gaete:2022lno}
M.~Bravo-Gaete, F.~F.~Santos and H.~Boschi-Filho,
``Shear viscosity from black holes in generalized scalar-tensor theories in arbitrary dimensions,''
Phys. Rev. D \textbf{106}, no.6, 066010 (2022),
doi:10.1103/PhysRevD.106.066010
[arXiv:2201.07961 [hep-th]].

\bibitem{Eisert:2008ur}
J.~Eisert, M.~Cramer and M.~B.~Plenio,
``Area laws for the entanglement entropy - a review,''
Rev. Mod. Phys. \textbf{82}, 277-306 (2010),
doi:10.1103/RevModPhys.82.277
[arXiv:0808.3773 [quant-ph]].

\bibitem{Berti:2009kk}
E.~Berti, V.~Cardoso and A.~O.~Starinets,
``Quasinormal modes of black holes and black branes,''
Class. Quant. Grav. \textbf{26}, 163001 (2009),
doi:10.1088/0264-9381/26/16/163001
[arXiv:0905.2975 [gr-qc]].


\bibitem{barrow}   J. D. Barrow, 
"The area of a rough black hole," 
Phys. Lett. B \textbf{808}, 135643 (2020),
doi:10.1016/j.physletb.2020.135643
[arXiv: 2004.09444 [gr-qc]].


\bibitem{Abdalla:2022yfr}
E.~Abdalla, G.~Franco Abell{\'a}n, A.~Aboubrahim, A.~Agnello, O.~Akarsu, Y.~Akrami, G.~Alestas, D.~Aloni, L.~Amendola and L.~A.~Anchordoqui, \textit{et al.}
``Cosmology intertwined: A review of the particle physics, astrophysics, and cosmology associated with the cosmological tensions and anomalies,''
JHEAp \textbf{34}, 49 (2022),
doi:10.1016/j.jheap.2022.04.002
[arXiv:2203.06142 [astro-ph.CO]].

\bibitem{Saridakis:2020zol}
E.~N.~Saridakis,
``Barrow holographic dark energy,''
Phys. Rev. D \textbf{102}, no.12, 123525 (2020),
doi:10.1103/PhysRevD.102.123525
[arXiv:2005.04115 [gr-qc]].

\bibitem{Anagnostopoulos:2020ctz}
F.~K.~Anagnostopoulos, S.~Basilakos and E.~N.~Saridakis,
``Observational constraints on Barrow holographic dark energy,''
Eur. Phys. J. C \textbf{80}, no.9, 826 (2020),
doi:10.1140/epjc/s10052-020-8360-5
[arXiv:2005.10302 [gr-qc]].

\bibitem{Nojiri:2022aof}
S.~Nojiri, S.~D.~Odintsov and V.~Faraoni,
``From nonextensive statistics and black hole entropy to the holographic dark universe,''
Phys. Rev. D \textbf{105}, no.4, 044042 (2022),
doi:10.1103/PhysRevD.105.044042
[arXiv:2201.02424 [gr-qc]].

\bibitem{Barrow:2020kug}
J.~D.~Barrow, S.~Basilakos and E.~N.~Saridakis,
``Big Bang Nucleosynthesis constraints on Barrow entropy,''
Phys. Lett. B \textbf{815}, 136134 (2021),
doi:10.1016/j.physletb.2021.136134
[arXiv:2010.00986 [gr-qc]].

\bibitem{Dimakis:2021gby}
N.~Dimakis, A.~Paliathanasis and T.~Christodoulakis,
``Quantum cosmology in f(Q) theory,''
Class. Quant. Grav. \textbf{38}, no.22, 225003 (2021),
doi:10.1088/1361-6382/ac2b09
[arXiv:2108.01970 [gr-qc]].

\bibitem{Moradpour:2020dfm}
H.~Moradpour, A.~H.~Ziaie and M.~Kord Zangeneh,
``Generalized entropies and corresponding holographic dark energy models,''
Eur. Phys. J. C \textbf{80}, no.8, 732 (2020),
doi:10.1140/epjc/s10052-020-8307-x
[arXiv:2005.06271 [gr-qc]].

\bibitem{Sheykhi:2021fwh}
A.~Sheykhi,
``Barrow Entropy Corrections to Friedmann Equations,''
Phys. Rev. D \textbf{103}, no.12, 123503 (2021),
doi:10.1103/PhysRevD.103.123503
[arXiv:2102.06550 [gr-qc]].

\bibitem{Nojiri:2021czz}
S.~Nojiri, S.~D.~Odintsov and V.~Faraoni,
``Area-law versus R{\'e}nyi and Tsallis black hole entropies,''
Phys. Rev. D \textbf{104}, no.8, 084030 (2021),
doi:10.1103/PhysRevD.104.084030
[arXiv:2109.05315 [gr-qc]].

\bibitem{Abreu:2020wbz}
E.~M.~C.~Abreu and J.~A.~Neto,
``Barrow black hole corrected-entropy model and Tsallis nonextensivity,''
Phys. Lett. B \textbf{810}, 135805 (2020),
doi:10.1016/j.physletb.2020.135805
[arXiv:2009.10133 [gr-qc]].


\bibitem{Abreu:2024uvp}
E.~M.~C.~Abreu,
``Barrow black hole variable parameter model connected to information theory,''
[arXiv:2402.15922 [gr-qc]].

\bibitem{Abreu:2025etn}
E.~M.~C.~Abreu and A.~C.~R.~Mendes,
``Modified Friedmann equations and dark energy analysis with a variable barrow black hole parameter,''
Mod. Phys. Lett. A \textbf{40}, no.34, 2550162 (2025),
doi:10.1142/S0217732325501627.

\bibitem{Basilakos:2023seo}
S.~Basilakos, A.~Lymperis, M.~Petronikolou and E.~N.~Saridakis,
``Barrow holographic dark energy with varying exponent,''
Nucl. Phys. B \textbf{1015}, 116904 (2025),
doi:10.1016/j.nuclphysb.2025.116904
[arXiv:2312.15767 [gr-qc]].

\bibitem{dolan} 
B.~P.~Dolan,
``Pressure and volume in the first law of black hole thermodynamics,''
Class. Quant. Grav. \textbf{28}, 235017 (2011),
doi:10.1088/0264-9381/28/23/235017
[arXiv:1106.6260 [gr-qc]]; 
``The cosmological constant and the black hole equation of state,''
Class. Quant. Grav. \textbf{28}, 125020 (2011),
doi:10.1088/0264-9381/28/12/125020
[arXiv:1008.5023 [gr-qc]].


\bibitem{Johnson:2014yja}
C.~V.~Johnson,
``Holographic Heat Engines,''
Class. Quant. Grav. \textbf{31}, 205002 (2014),
doi:10.1088/0264-9381/31/20/205002
[arXiv:1404.5982 [hep-th]].

\bibitem{Caceres:2015vsa}
E.~Caceres, P.~H.~Nguyen and J.~F.~Pedraza,
``Holographic entanglement entropy and the extended phase structure of STU black holes,''
JHEP \textbf{09}, 184 (2015),
doi:10.1007/JHEP09(2015)184
[arXiv:1507.06069 [hep-th]].

\bibitem{Johnson:2015ekr}
C.~V.~Johnson,
``Gauss{\textendash}Bonnet black holes and holographic heat engines beyond large $N$,''
Class. Quant. Grav. \textbf{33}, no.21, 215009 (2016),
doi:10.1088/0264-9381/33/21/215009
[arXiv:1511.08782 [hep-th]].

\bibitem{Johnson:2015fva}
C.~V.~Johnson,
``Born{\textendash}Infeld AdS black holes as heat engines,''
Class. Quant. Grav. \textbf{33}, no.13, 135001 (2016)
doi:10.1088/0264-9381/33/13/135001
[arXiv:1512.01746 [hep-th]].


\bibitem{Hendi:2017bys}
S.~H.~Hendi, B.~Eslam Panah, S.~Panahiyan, H.~Liu and X.~H.~Meng,
``Black holes in massive gravity as heat engines,''
Phys. Lett. B \textbf{781}, 40-47 (2018),
doi:10.1016/j.physletb.2018.03.072
[arXiv:1707.02231 [hep-th]].

\bibitem{Johnson:2016pfa}
C.~V.~Johnson,
``An Exact Efficiency Formula for Holographic Heat Engines,''
Entropy \textbf{18}, 120 (2016),
doi:10.3390/e18040120
[arXiv:1602.02838 [hep-th]].

\bibitem{Chandrasekhar:2016lbd}
B.~Chandrasekhar and P.~K.~Yerra,
``Heat engines for dilatonic Born{\textendash}Infeld black holes,''
Eur. Phys. J. C \textbf{77}, no.8, 534 (2017),
doi:10.1140/epjc/s10052-017-5076-2
[arXiv:1606.03223 [hep-th]].

\bibitem{Wei:2017vqs}
S.~W.~Wei and Y.~X.~Liu,
``Charged AdS black hole heat engines,''
Nucl. Phys. \textbf{B}, 114700 (2019),
doi:10.1016/j.nuclphysb.2019.114700
[arXiv:1708.08176 [gr-qc]].

\bibitem{Maldacena:1997re}
J.~M.~Maldacena,
``The Large N limit of superconformal field theories and supergravity,''
Adv. Theor. Math. Phys. \textbf{2}, 231-252 (1998),
doi:10.1023/A:1026654312961
[arXiv:hep-th/9711200 [hep-th]].

\bibitem{Witten:1998zw}
E.~Witten,
``Anti-de Sitter space, thermal phase transition, and confinement in gauge theories,''
Adv. Theor. Math. Phys. \textbf{2}, 505 (1998),
doi:10.4310/ATMP.1998.v2.n3.a3
[arXiv:hep-th/9803131 [hep-th]].

\bibitem{Chamblin:1999tk}
A.~Chamblin, R.~Emparan, C.~V.~Johnson and R.~C.~Myers,
``Charged AdS black holes and catastrophic holography,''
Phys. Rev. D \textbf{60}, 064018 (1999),
doi:10.1103/PhysRevD.60.064018
[arXiv:hep-th/9902170 [hep-th]].
\bibitem{Chamblin:1999hg}
A.~Chamblin, R.~Emparan, C.~V.~Johnson and R.~C.~Myers,
``Holography, thermodynamics and fluctuations of charged AdS black holes,''
Phys. Rev. D \textbf{60}, 104026 (1999),
doi:10.1103/PhysRevD.60.104026
[arXiv:hep-th/9904197 [hep-th]].
\bibitem{Kastor:2009wy}
D.~Kastor, S.~Ray and J.~Traschen,
``Enthalpy and the Mechanics of AdS Black Holes,''
Class. Quant. Grav. \textbf{26}, 195011 (2009), 
doi:10.1088/0264-9381/26/19/195011
[arXiv:0904.2765 [hep-th]].

\bibitem{Kubiznak:2012wp}
D.~Kubiznak and R.~B.~Mann,
``P-V criticality of charged AdS black holes,''
JHEP \textbf{07}, 033 (2012), 
doi:10.1007/JHEP07(2012)033
[arXiv:1205.0559 [hep-th]].


\bibitem{Lobo:2022eyr}
I.~P.~Lobo, J.~P.~Morais Gra{\c{c}}a, E.~Folco Capossoli and H.~Boschi-Filho,
``A varying gravitational constant map in asymptotically AdS black hole thermodynamics,''
Phys. Lett. B \textbf{835}, 137559 (2022)
doi:10.1016/j.physletb.2022.137559
[arXiv:2206.13664 [hep-th]].

\bibitem{Okcu:2016tgt}
\"O.~\"Okc\"u and E.~Ayd\i{}ner,
``Joule\textendash{}Thomson expansion of the charged AdS black holes,''
Eur. Phys. J. C \textbf{77}, 24 (2017),
doi:10.1140/epjc/s10052-017-4598-y
[arXiv:1611.06327 [gr-qc]].



\bibitem{Okcu:2017qgo}
\"O.~\"Okc\"u and E.~Ayd\i{}ner,
`Joule\textendash{}Thomson expansion of Kerr\textendash{}AdS black holes,''
Eur. Phys. J. C \textbf{78}, 123 (2018),
doi:10.1140/epjc/s10052-018-5602-x
[arXiv:1709.06426 [gr-qc]].

\bibitem{Mo:2018rgq}
J.~X.~Mo, G.~Q.~Li, S.~Q.~Lan and X.~B.~Xu,
``Joule-Thomson expansion of $d$-dimensional charged AdS black holes,''
Phys. Rev. D \textbf{98}, 124032 (2018),
doi:10.1103/PhysRevD.98.124032
[arXiv:1804.02650 [gr-qc]].

\bibitem{Lan:2018nnp}
S.~Q.~Lan,
``Joule-Thomson expansion of charged Gauss-Bonnet black holes in AdS space,''
Phys. Rev. D \textbf{98}, 084014 (2018),
doi:10.1103/PhysRevD.98.084014
[arXiv:1805.05817 [gr-qc]].

\bibitem{Mo:2018qkt}
J.~X.~Mo and G.~Q.~Li,
``Effects of Lovelock gravity on the Joule\textendash{}Thomson expansion,''
Class. Quant. Grav. \textbf{37}, 045009 (2020),
doi:10.1088/1361-6382/ab60b9
[arXiv:1805.04327 [gr-qc]].

\bibitem{Cisterna:2018jqg}
A.~Cisterna, S.~Q.~Hu and X.~M.~Kuang,
``Joule-Thomson expansion in AdS black holes with momentum relaxation,''
Phys. Lett. B \textbf{797}, 134883 (2019), 
doi:10.1016/j.physletb.2019.134883
[arXiv:1808.07392 [gr-qc]].

\bibitem{Rizwan:2018mpy}
A.~Rizwan C.L., N.~Kumara A., D.~Vaid and K.~M.~Ajith,
``Joule-Thomson expansion in AdS black hole with a global monopole,''
Int. J. Mod. Phys. A \textbf{33}, 1850210 (2019),
doi:10.1142/S0217751X1850210X
[arXiv:1805.11053 [gr-qc]].

\bibitem{Chabab:2018zix}
M.~Chabab, H.~El Moumni, S.~Iraoui, K.~Masmar and S.~Zhizeh,
``Joule-Thomson Expansion of RN-AdS Black Holes in $f(R)$ gravity,''
LHEP \textbf{02}, 05 (2018), 
doi:10.31526/LHEP.2.2018.02
[arXiv:1804.10042 [gr-qc]].

\bibitem{Yekta:2019wmt}
D.~Mahdavian Yekta, A.~Hadikhani and \"O.~\"Okc\"u,
``Joule-Thomson expansion of charged AdS black holes in Rainbow gravity,''
Phys. Lett. B \textbf{795}, 521 (2019), 
doi:10.1016/j.physletb.2019.06.049
[arXiv:1905.03057 [hep-th]].

\bibitem{Li:2019jcd}
C.~Li, P.~He, P.~Li and J.~B.~Deng,
``Joule-Thomson expansion of the Bardeen-AdS black holes,''
Gen. Rel. Grav. \textbf{52}, 50 (2020),
doi:10.1007/s10714-020-02704-z
[arXiv:1904.09548 [gr-qc]].

\bibitem{Nam:2019zyk}
C.~H.~Nam,
``Heat engine efficiency and Joule\textendash{}Thomson expansion of nonlinear charged AdS black hole in massive gravity,''
Gen. Rel. Grav. \textbf{53}, 30 (2021),
doi:10.1007/s10714-021-02787-2
[arXiv:1906.05557 [gr-qc]].

\bibitem{K.:2020rzl}
K.~V.~Rajani, C.~L.~A.~Rizwan, A.~Naveena Kumara, M.~S.~Ali and D.~Vaid,
``Joule{\textendash}Thomson expansion of regular Bardeen AdS black hole surrounded by static anisotropic matter field,''
Phys. Dark Univ. \textbf{32}, 100825 (2021)
doi:10.1016/j.dark.2021.100825
[arXiv:2002.03634 [gr-qc]].


\bibitem{Bi:2020vcg}
S.~Bi, M.~Du, J.~Tao and F.~Yao,
``Joule-Thomson expansion of Born-Infeld AdS black holes,''
Chin. Phys. C \textbf{45}, no.2, 025109 (2021),
doi:10.1088/1674-1137/abcf23
[arXiv:2006.08920 [gr-qc]].


\bibitem{boschiJT1}   J. P. M. Gra\c{c}a, E. F. Capossoli, H. Boschi-Filho and I. P. Lobo, 
"Joule-Thomson expansion for quantum corrected AdS-Reissner-Nordström black hole in a
Kiselev spacetime," 
Phys. Rev. D 107, 024045 (2023), 
https://doi.org/10.1103/PhysRevD.107.024045
[arXiv: 2105.04689 [gr-qc]].

\bibitem{boschiJT2}
J. P. M. Gra\c{c}a, E. F. Capossoli and H. Boschi-Filho, 
"Joule-Thomson expansion for noncommutative uncharged black holes," 
EPL 135, 41002 (2021),
doi:10.1209/0295-5075/ac24c1
[arXiv: 2107.05781[hep-th]].

\bibitem{Silva:2021qtw}
G.~V.~Silva, V.~B.~Bezerra, J.~P.~M.~Gra{\c{c}}a and I.~P.~Lobo,
``Joule{\textendash}Thomson expansion in charged AdS black hole surrounded by a cosmological fluid in Rainbow Gravity,''
Mod. Phys. Lett. A \textbf{36}, no.40, 2150278 (2021),
doi:10.1142/S0217732321502783.

\bibitem{Yin:2021akt}
R.~Yin, J.~Liang and B.~Mu,
``Joule{\textendash}Thomson expansion of Reissner{\textendash}Nordstr{\"o}m-Anti-de Sitter black holes with cloud of strings and quintessence,''
Phys. Dark Univ. \textbf{34}, 100884 (2021),
doi:10.1016/j.dark.2021.100884
[arXiv:2105.09173 [gr-qc]].

\bibitem{Biswas:2021uop}
A.~Biswas,
``Joule-Thomson expansion of AdS black holes in Einstein Power-Yang-mills gravity,''
Phys. Scripta \textbf{96}, no.12, 125310 (2021),
doi:10.1088/1402-4896/ac2b42
[arXiv:2106.11066 [gr-qc]].

\bibitem{Xing:2021gpn}
J.~T.~Xing, Y.~Meng and X.~M.~Kuang,
``Joule-Thomson expansion for hairy black holes,''
Phys. Lett. B \textbf{820}, 136604 (2021),
doi:10.1016/j.physletb.2021.136604.

\bibitem{Cao:2022hmd}
Y.~Cao, H.~Feng, J.~Tao and Y.~Xue,
``Black holes in a cavity: Heat engine and Joule-Thomson expansion,''
Gen. Rel. Grav. \textbf{54}, no.9, 105 (2022),
doi:10.1007/s10714-022-02990-9
[arXiv:2201.07584 [gr-qc]].

\bibitem{Mondal:2022ozv}
D.~Mondal and U.~Debnath,
``Thermodynamics of Power{\textendash}Maxwell charged AdS black holes with quintessence in Rastall gravity: Heat engine,''
Int. J. Mod. Phys. A \textbf{37}, no.10, 2250058 (2022),
doi:10.1142/S0217751X22500580.

\bibitem{Bai:2022yno}
Y.~Y.~Bai,
``Joule{\textendash}Thomson expansion of Einstein{\textendash}Skyrmion black holes,''
Can. J. Phys. \textbf{101}, no.1, 1 (2023),
doi:10.1139/cjp-2022-0062.

\bibitem{Masmar:2023qol}
K.~Masmar,
``Joule{\textendash}Thomson expansion for a nonlinearly charged Anti-de Sitter black hole,''
Int. J. Geom. Meth. Mod. Phys. \textbf{20}, no.05, 2350080 (2023),
doi:10.1142/S0219887823500809.

\bibitem{Li:2023mql}
N.~Li, J.~Y.~Li and B.~Y.~Su,
``The Joule{\textendash}Thomson and Joule{\textendash}Thomson-Like Effects of the Black Holes in a Cavity,''
Fortsch. Phys. \textbf{71}, no.2-3, 2200166 (2023),
doi:10.1002/prop.202200166
[arXiv:2306.15959 [gr-qc]].

\bibitem{Alipour:2024xxs}
M.~R.~Alipour, S.~Noori Gashti, M.~A.~S.~Afshar and J.~Sadeghi,
``Cooling and heating regions of Joule-Thomson expansion for AdS black holes: Einstein-Maxwell-power-Yang-Mills and Kerr Sen black holes,''
Gen. Rel. Grav. \textbf{57}, no.3, 61 (2025),
doi:10.1007/s10714-025-03393-2
[arXiv:2402.02257 [hep-th]].

\bibitem{Yasir:2024gza}
M.~Yasir, T.~Xia, A.~Ditta, D.~Arora, F.~Atamurotov, A.~Mahmood and O.~Egamberdiev,
``Joule{\textendash}Thomson expansion of Bardeen black hole with a cloud of strings,''
Int. J. Mod. Phys. A \textbf{39}, no.11n12, 2450046 (2024),
doi:10.1142/S0217751X24500465.

\bibitem{Javed:2024ohv}
F.~Javed, A.~Waseem, P.~Channuie, G.~Mustafa, T.~Muhammad and E.~G{\"u}dekli,
``Particle dynamics and Joule{\textendash}Thomson expansion of phantom anti-de Sitter black hole stability and thermal fluctuations in massive gravity,''
Phys. Dark Univ. \textbf{47}, 101766 (2025),
doi:10.1016/j.dark.2024.101766.


\bibitem{Dai:2025fxl}
E.~Dai, F.~Javed, A.~Waseem, M.~Alosaimi and R.~M.~Zulqarnain,
``Thermodynamic insights into Joule{\textendash}Thomson expansion, particle dynamics, and emission energy of AdS black holes in Horndeski theory,''
Phys. Dark Univ. \textbf{49}, 102014 (2025),
doi:10.1016/j.dark.2025.102014.


\bibitem{Reif} 
F. Reif, Fundamentals of statistical and thermal physics, McGraw-Hill, 1965. 


\bibitem{Javed:2024nnt}
F.~Javed, G.~Mustafa, G.~Fatima, S.~K.~Maurya, M.~H.~Alshehri and I.~Mubeen,
``Joule-Thomson expansion for charged-AdS black hole with nonlinear electrodynamics and thermal fluctuations by using Barrow entropy,''
JHEAp \textbf{44}, 60 (2024),
doi:10.1016/j.jheap.2024.09.003.

\bibitem{Yasir:2024oah}
M.~Yasir, X.~Tiecheng, S.~Chaudhary and A.~B.~Jumah,
``Quantum-gravitational effects on Joule-Thomson expansion and black hole shadows in F(R) gravity with barrow entropy corrections,''
JHEAp \textbf{44}, 356 (2024),
doi:10.1016/j.jheap.2024.10.006.

\bibitem{Javed:2025dit}
F.~Javed, M.~Zeeshan Gul, O.~Donmez, T.~Naseer and M.~H.~Alshehri,
``Joule-Thomson expansion with Barrow entropy and particle dynamics of charged Rastall-AdS black hole,'' 
Nucl. Phys. B \textbf{1018}, 117001 (2025),
doi:10.1016/j.nuclphysb.2025.117001.

\bibitem{Ladghami:2024yjn}
Y.~Ladghami, A.~Bargach, A.~Bouali, T.~Ouali and G.~Mustafa,
``Spacetime foam effects on charged AdS black hole thermodynamics,'' 
Nucl. Phys. B \textbf{1018}, 117015 (2025),
doi:10.1016/j.nuclphysb.2025.117015
[arXiv:2411.06271 [hep-th]].

\bibitem{Biswas:2025jxb}
R.~Biswas and S.~Pal,
``Einstein Maxwell Scalar Black Hole: Thermodynamic Properties with Logarithmic Barrow Entropy,''
[arXiv:2505.17172 [gr-qc]].







\bibitem{Kazakov:1993ha}
D.~I.~Kazakov and S.~N.~Solodukhin,
``On Quantum deformation of the Schwarzschild solution,''
Nucl. Phys. B \textbf{429}, 153 (1994), 
doi:10.1016/S0550-3213(94)80045-6
[arXiv:hep-th/9310150 [hep-th]].

\bibitem{Lobo:2019put}
V.~B.~Bezerra, I.~P.~Lobo, J.~P.~Morais Gra\c{c}a and L.~C.~N.~Santos,
``Effects of quantum corrections on the criticality and efficiency of black holes surrounded by a perfect fluid,''
Eur. Phys. J. C \textbf{79}, no.11, 949 (2019),
doi:10.1140/epjc/s10052-019-7482-0
[arXiv:1908.08140 [gr-qc]].
\bibitem{Kiselev:2002dx}
V.~V.~Kiselev,
``Quintessence and black holes,''
Class. Quant. Grav. \textbf{20}, 1187 (2003),
doi:10.1088/0264-9381/20/6/310
[arXiv:gr-qc/0210040 [gr-qc]].
\bibitem{Caldwell:1999ew}
R.~R.~Caldwell,
``A Phantom menace?,''
Phys. Lett. B \textbf{545}, 23 (2002),
doi:10.1016/S0370-2693(02)02589-3
[arXiv:astro-ph/9908168 [astro-ph]].



\bibitem{Visser:2019brz}
M.~Visser,
``The Kiselev black hole is neither perfect fluid, nor is it quintessence,''
Class. Quant. Grav. \textbf{37}, 045001 (2020),
doi:10.1088/1361-6382/ab60b8
[arXiv:1908.11058 [gr-qc]].


\bibitem{Shahjalal:2019pqb}
M.~Shahjalal,
``Thermodynamics of quantum-corrected Schwarzschild black hole surrounded by quintessence,''
Nucl. Phys. B \textbf{940}, 63 (2019),
doi:10.1016/j.nuclphysb.2019.01.009.





\bibitem{Ji:2023epw}
Y.~Ji, J.~Chen, G.~L.~Schumacher, G.~G.~T.~Assump{\c{c}}{\~a}o, S.~Huang, F.~J.~Vivanco and N.~Navon,
``Observation of the Fermionic Joule-Thomson Effect,''
Phys. Rev. Lett. \textbf{132}, no.15, 153402 (2024),
doi:10.1103/PhysRevLett.132.153402
[arXiv:2305.16320 [cond-mat.quant-gas]].



\end{thebibliography}
\end{document}